\RequirePackage{fix-cm}
\documentclass[twocolumn,epjc3]{svjour3}

\RequirePackage{color,graphicx,url}

\journalname{Eur. Phys. J. A}

\usepackage{amsfonts,amsmath,amssymb,bm,xspace}

\graphicspath{{./figures/}}

\usepackage[colorlinks=true]{hyperref}  

\graphicspath{{figures/}} 
\usepackage{multirow}

\usepackage{academicons} 
\newcommand{\orcid}[1]{\href{https://orcid.org/#1}{\textcolor[HTML]{A6CE39}{\aiOrcid}}}
\usepackage{xcolor}

\usepackage{bm}
\usepackage{enumitem}

\usepackage{nccmath}

\usepackage{xfrac}

\newcommand{\sat}{\mathrm{sat}}
\newcommand{\sym}{\mathrm{sym}}

\renewcommand{\L}{\mathcal{L}}
\newcommand{\psib}{\bar{\psi}}

\hypersetup{%
    pdfsubject=Paper,
    pdfkeywords={nuclear physics} {MBPT} {chiral EFT} {asymmetric nuclear matter}
    unicode=true,
    breaklinks=true,
     colorlinks   = true,
     linkcolor = blue,
     citecolor = blue,
     menucolor = blue,
     urlcolor = blue
}

\begin{document}

\title{Relativistic Hartree-Fock Chiral Lagrangians with confinement, nucleon finite size and short-range effects}

\author{Mohamad Chamseddine\thanksref{addr1}{\includegraphics[scale=0.3]{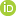}} 
\and
J\'er\^ome Margueron\thanksref{addr1}\href{https://orcid.org/0000-0001-8743-3092}{\includegraphics[scale=0.3]{figures/orcid.png}}
\and
Guy Chanfray\thanksref{addr1}\href{https://orcid.org/0000-0002-3593-9507}{\includegraphics[scale=0.3]{figures/orcid.png}}  
\and
Hubert Hansen\thanksref{addr1}\href{https://orcid.org/0000-0001-8879-3612}{\includegraphics[scale=0.3]{figures/orcid.png}} 
\and
Rahul Somasundaram\thanksref{addr2,addr3}\href{https://orcid.org/0000-0003-0427-3893}{\includegraphics[scale=0.3]{figures/orcid.png}} 
}

\institute{Univ Lyon, Univ Claude Bernard Lyon 1, CNRS/IN2P3, IP2I Lyon, UMR 5822, F-69622, Villeurbanne, France \label{addr1}
\and
Theoretical Division, Los Alamos National Laboratory, Los Alamos, New Mexico 87545, USA\label{addr2}
\and
Department of Physics, Syracuse University, Syracuse, NY 13244, USA\label{addr3}
}






\date{Draft version:\today, Received: date / Accepted: date}

\maketitle

\abstract{A relativistic Hartree-Fock Lagrangian including a chiral potential and nucleon polarisation is investigated in hopes of providing a better description of dense nuclear matter. We fully consider the contribution of the exchange Fock term to the energy and the self-energies, and in addition we investigate the nucleon's compositeness and finite size effects (confinement and form factors) and short range correlations modeled by a Jastrow ansatz. These effects are added step by step, such that their impact on the dense matter properties can be analysed in details. The parameters of the model are adjusted to reproduce fundamental properties related to the QCD theory at low energy, such as the chiral symmetry breaking, nucleon's quark substructure and Lattice-QCD predictions, as well as two empirical properties at saturation: the binding energy and the density. All other empirical parameters, e.g., symmetry energy and its slope, incompressibility modulus, effective mass, as well as spin-isospin Landau-Midgal parameter are predictions of the models and can be used to evaluate the gain of the different approximation schemes in describing nuclear properties. Bayesian statistics is employed in order to propagate parameter uncertainties into predictions for the nuclear matter properties. We show that the splitting of the effective Landau mass is largely influenced by the value of the $\rho^T$ coupling, and we show that the fit to the symmetry energy, which induces an increase of the coupling constant $g_\rho$ by about 20-25\% compared to the case where it is fixed by the quark model, provides a very good EoS compatible with the present nuclear physics knowledge.}
 

\maketitle


\section{Introduction}
\label{sec:intro}

The theory of the strong interaction, the so-called quantum chromo-dynamics (QCD), has been formulated in the seventies~\cite{QCD1973} and provides good description of high energy processes which can be found at high temperatures or high densities. In these regimes, QCD is perturbative, which explains why its direct application is possible. However, at low energies, such as low-temperature or low-density encountered in the atomic nucleus, this theory is non-perturbative. It is reflected in different phenomena such as the spontaneous breaking of symmetries that gives a non-trivial structure to the QCD vacuum as well as the color confinement, which is under investigation, for instance, by the CBM experiment at FAIR~\cite{CBM2017}. In the low energy regime where perturbative approaches do not apply directly, several effective models capturing the  symmetries of the theory has been suggested, such as, the Nambu-Jona-Lasinio (NJL) model~\cite{Nambu:1961fr} or the effective chiral approach~\cite{Weinberg}, also known as Chiral Effective Field Theory ($\chi$-EFT). The latter approach, is well suited to explore low-density nuclear matter and breaks down as the density exceeds 
$n_{\sat}$ to $2n_{\sat}$~\cite{Tews:2018kmu}, where $n_{\sat}$ is the nuclear saturation density ($n_{\sat}\approx 0.155\pm 0.005$~fm$^{-3}$~\cite{Margueron2018}). The exploration of the densest phase of nuclear matter therefore requires extrapolations such as the agnostic ones proposed in Refs.~\cite{Hebeler2013,Somasundaram:2021clp} for instance.

The important features of QCD at low-energy are the following: the degrees of freedom are nucleons and mesons, instead of quarks and gluons, the vacuum is non-trivial due to the spontaneous breaking of the chiral symmetry, and the fields are in, or close to, a relativistic regime. In the present paper, we investigate a chiral Lagrangian for nucleons and mesons, where the spontaneous breaking of the chiral symmetry gives rise to a scalar chiral potential, and where the quark substructure is reflected in the polarisability of the nucleons at finite density. We additionally consider the relativistic dynamics of the fields in the present approach.

In this work, we consider the mean-field approach, and beyond, in treating the nuclear interaction. Since we are dealing with an N-body problem, this technique consists of taking nucleons that are bathing in a background classical "mean field", scalar or vector, and this is known as the Hartree level. Going beyond that, the quantum fluctuations of these fields should be taken into account in what is known as the Fock terms, which in a quantum field theory would correspond to an interaction mediated by mesons, hence why it is common to see it described as meson exchange. The Hartree level was previously studied in
\textit{Somasundaram et al.}~\cite{Rahul2022} for a relativistic mean-field model where spontaneous chiral symmetry breaking and confinement effects are incorporated (RMF-CC model). This model is based on two important concepts: the identification of the nuclear physics $\sigma$ meson, introduced in relativistic mean fields approaches, with the radial fluctuation of the chiral quark condensate as proposed in Ref.~\cite{Chanfray2001}, and the inclusion of a nucleon "polarisation" by the nuclear environment as in Ref.~\cite{Chanfray2005}. As mentioned earlier, this study was done at the Hartree level, and we now incorporate the Fock term. The Fock term is indeed instrumental in order to incorporate the missing contribution of the pion meson for instance. In the present study, we consider symmetric matter (SM) and neutron matter (NM), and we analyze the impact of the form factor (FF) at the interaction vertices and the inclusion of short range correlation (SRC) due to the strong repulsion of the potential at short distance.

The study is organized in the following way: In Sec.~\ref{sec:Formalism} we briefly describe the content of our relativistic mean field Lagrangian. Sec.~\ref{sec:caseA} is devoted to the formalism of the pure relativistic Hartree-Fock approach (called model A in the following) then sec.~\ref{sec:Improvment} presents various improvements of the Fock term, such as the widely used one in nuclear physics where the contact contribution of the spin-isospin $\pi$ and $\rho$ exchange are fully removed by the so-called Orsay prescription~\cite{Bouyssy_1987} (model B), or other approaches where the finite size effects (model C) and SRC (model D) are considered. The SRC is implemented through the well-known Jastrow ansatz as used in Oset-Toki-Weise paper~\cite{OSET1982281}. We then present our results in Sec.~\ref{sec:Results}, where we perform sensitivity analyses as well as detailed study based on the Bayesian approach. We also show the important contribution of the $\rho^T$ coupling to the question of the splitting of the effective Landau mass in neutron matter. Our conclusions are given in Sec.~\ref{sec:conclusion}.

\section{The relativistic mean field Lagrangian}
\label{sec:Formalism}

\subsection{Formalism}

Considering only the lightest $u$ and $d$ quarks and the flavor number $N_f=2$, the chiral fields associated to the fluctuations of the quark condensate $\langle \bar q q\rangle$ resulting from chiral symmetry breaking are usually parameterized in term of a $\rm{SU}(2)$ matrix $M$ as:
\begin{equation}
M=\sigma + i\vec{\tau}\cdot\vec{\phi}\equiv S\, U
\label{REPRES}
\end{equation}
with $S = s + f_\pi$ and $U=e^{i\,{\vec{\tau}\cdot\vec{\pi}}/{f_\pi}}$.
The scalar field $\sigma$ ($S$) and pseudo-scalar fields $\vec{\phi}$ ($\vec{\pi}$) written in Cartesian (polar) coordinates  appear as the dynamical degrees of freedom. As stated in the introduction, it is necessary to clarify the connection between the nuclear physics sigma meson of the Walecka model (let us call it $\sigma_W$ from now on) at the origin of the nuclear binding with a chiral field~\eqref{REPRES}. For instance, one may be tempted to identify $\sigma_W$ with the scalar field $\sigma$ in Cartesian coordinates. It is however forbidden by chiral constraints and this point has been first addressed by Birse~\cite{Birse94}: it would lead to the presence of terms of order $m_\pi$ in the NN interaction which is not allowed. 

In this study, we follow Ref.~\cite{Chanfray2001} and identify $\sigma_W$ with the chiral invariant $s$ ($=S-f_\pi$) field associated with the radial fluctuation of the chiral condensate $S$ around the \emph{chiral radius} $f_\pi$, in polar coordinates. It formally consists of promoting the chiral invariant scalar field $s$ and the pion field $\vec{\pi}$ appearing in the matrix $M$ in Eq.~\eqref{REPRES}
to effective degrees of freedom. This was originally formulated in the framework of the linear sigma model (L$\sigma$M)~\cite{Chanfray2001} but an explicit construction using a bosonization technique of the chiral effective potential can be done within the NJL model~\cite{Chanfray2011} where the linear sigma model potential is recovered through a second order expansion in  $S^2-f^2_\pi$ of the constituent quark Dirac sea energy.
This proposal, which gives a plausible answer to the long standing problem of the chiral status of Walecka theories, has also the merit of respecting all the desired chiral constraints~\cite{Birse94}. In particular the correspondence $s\equiv \sigma_W$ generates a coupling of the scalar field to the derivatives of the pion field, as expected in the physical world. Hence the radial mode decouples from low-energy pions whose dynamics is governed by chiral perturbation theory. A detailed discussion of this sometimes subtle topic is given in \cite{Chanfray2001,Chanfray2006,Hebeler2013} and in the following we employ the notation $s$ for the scalar field to avoid confusion with the meson $\sigma$.

After the bosonisation of the quark fields and the introduction of a confining potential~\cite{Chanfray2011}, the new degrees of freedom are the nucleons and the mesons.
The relativistic Lagrangian can generically be written as the sum of a kinetic fermionic term,
\begin{equation}
\label{eq:L_kinetic}
\L_\psi = \psib \left( i \gamma^{\mu} -M_N \right)\partial_{\mu} \psi \, ,
\end{equation}
where the field $\psi$ represents the nucleon spinor, and of meson-nucleon terms,
\begin{equation}
\L_{m} = \L_{s} + \L_{\omega} + \L_{\rho} + \L_{\delta} + \L_{\pi} \, ,
\label{eq:L_mesons}
\end{equation}
collecting all mesonic contributions considered in a given model. Using notation of Ref.~\cite{Massot2008} these can be enumerated as,
\begin{align}
\label{eq:L_meson}
\L_s =& \big(M_N - M_N(s)\big) \bar{\psi} \psi - v(s) + \frac{1}{2} \partial^\mu s \partial_\mu s\, , \nonumber \\
\L_\omega =& -g_\omega \omega_\mu \bar{\psi} \gamma^\mu \psi + \frac{1}{2} m_\omega^2 \omega^\mu \omega_\mu - \frac{1}{4} F^{\mu \nu} F_{\mu \nu}\, , \nonumber \\
\L_\rho =& -g_\rho \rho_{a \mu} \bar{\psi} \gamma^\mu \tau_a \psi + g_\rho \frac{\kappa_\rho}{2 M_N} \partial_\nu \rho_{a \mu} \bar{\psi} \sigma^{\mu \nu} \tau_a \psi \nonumber \\
&+ \frac{1}{2} m_\rho^2 \rho_{a \mu} \rho_a^\mu - \frac{1}{4} G_a^{\mu \nu} G_{a \mu \nu}\, , \\
\L_\delta =& -g_\delta \delta_a \bar{\psi} \tau_a \psi - \frac{1}{2} m_\delta \delta_a \delta_a + \frac{1}{2} \partial^\mu \delta_a \partial_\mu \delta_a\, , \nonumber \\
\L_\pi =& \frac{g_A}{2 f_\pi} \partial_\mu \varphi_{\pi a} \bar{\psi} \gamma^\mu \gamma^5 \tau_a \psi - \frac{1}{2} m_\pi^2 \varphi_{\pi a} \varphi_{\pi a} \nonumber \\
&+ \frac{1}{2} \partial^\mu \varphi_{\pi a} \partial_\mu \varphi_{\pi a}\, , \nonumber
\end{align}
where the symbols have their usual meaning. Historically, the pion coupling was defined as $\tilde{f}_\pi/m_\pi$, which is replaced by $g_A/(2f_\pi)$ considering $\tilde{f}_\pi=m_\pi g_A/(2 f_\pi)$, where $g_A$ is the axial coupling constant and $f_\pi$ is the pion decay constant.
In Eq.~\eqref{eq:L_meson}, two quantities are of particular interest to us, the scalar potential $v(s)$ and the $s$-field dependent nucleon mass $M_N(s)$. We study the relativistic Hartree-Fock approach including chiral symmetry breaking through the chiral potential $v(s)$ and confinement (RHF-CC), in light of the RMF-CC in \cite{Rahul2022}, hereafter called (RH$_\mathrm{CC}$). The leading order effect of quark confinement is incorporated from its contribution to the nucleon polarisability associated with the readjustment of the quark wave function in the nuclear scalar field. We extend this model to include the finite-size effect through form factors (FF) as well as short-range correlations (SRC) from the Jastrow ansatz.

The effective chiral potential $v(s)$ in the scalar field $s$ has a typical Mexican hat shape which breaks chiral symmetry. We consider in this study the expression provided by the L$\sigma$M,
\begin{eqnarray}
v(s)&=& \frac{\lambda}{4}\big((f_\pi+s)^2-v^2\big)^2-f_\pi m_\pi^2s\nonumber\\
&\equiv &\frac{m^{2}_s}{2}s ^{2}+
\frac{m^{2}_s-m^{2}_{\pi }}{2f_\pi}s^3+
\frac{m^{2}_s-m^{2}_{\pi }}{8f_\pi^2}s^4.
\end{eqnarray}
More details on this effective chiral potential can be found, for instance, in Ref~\cite{Rahul2022}.

\begin{table*}[t]
\tabcolsep=0.5cm
\def\arraystretch{1.5}
\caption{\label{tab:parameters}%
Model parameters (masses and coupling constants) which are fixed to be constant in the present analysis. Note however that the nucleon-$\rho$ vector coupling is fixed from two prescriptions: i) to the quark model imposing that $g_{\rho} = g_{\omega}/3$, or ii) to reproduce the symmetry energy $E_\sym$. In the latter case, $g_\rho$ becomes a variable parameter in the fit protocol.}
\begin{tabular}{ccccccccc}
\hline
$M_N$ & $m_{\rho}$ & $m_{\delta}$ & $m_{\omega}$ & $m_{\pi}$ & $g_{\rho}$ & $g_{\delta}$ & $g_A$ & $f_{\pi}$ \\
MeV & MeV & MeV & MeV & MeV & & & & MeV \\
\hline
938.9 & 779.0 & 984.7 & 783.0 & 139.6 & i) quark model: $g_{\omega} / 3$ & 1 & 1.25 & 94.0\\
938.9 & 779.0 & 984.7 & 783.0 & 139.6 & ii) Fit parameter & 1 & 1.25 & 94.0\\
\hline
\end{tabular}
\end{table*}

\begin{table}[t]
\tabcolsep=0.66cm
\def\arraystretch{1.5}
\caption{\label{tab:rhotensor}%
We consider three cases for $\kappa_{\rho}$, the coupling constant of the $\rho_T$: no $\rho_T$ with $\kappa_{\rho}$=0.0 (for reference), weak $\rho_T$ with $\kappa_{\rho}$=3.7 suggested by the Vector Dominance Model (VDM) \cite{Bhaduri1988}, and strong $\rho_T$ with $\kappa_{\rho}$=6.6 suggested by scattering data~\cite{HOHLER1975210}. }
\begin{tabular}{cccc}
\hline
$\rho_T$ model & & Ref. & $\kappa_{\rho}$ \\
\hline
no $\rho_T$ & NRT & & 0.0\\
weak $\rho_T$ & WRT  & \cite{Bhaduri1988} & 3.7 \\
strong $\rho_T$ & SRT &\cite{HOHLER1975210} & 6.6 \\
\hline
\end{tabular}
\end{table}

In the presence of the nuclear scalar field, the nucleon mass is modified according to:
\begin{equation}
\label{eq:nucleon_mass}
M_N(s)=M_N+g_s s+\frac{1}{2}\kappa_\mathrm{NS}\left(s^2+\frac{s^3}{3f_\pi}\right).
\end{equation}
Here we do not take for the first order response, namely the scalar coupling constant, the value $g_S=M_N/f_\pi$ of the L$\sigma M$, but take it as a parameter possibly fixed by an underlying nucleon  model. The nucleon polarisability $\kappa_\mathrm{NS}$, incorporates the effect of the nucleon response, i.e., the central ingredient of the quark-meson coupling model (QMC) introduced in the original pioneering work of P. Guichon \cite{Guichon1988}. As in \cite{Chanfray2005} we include in practice  a scalar field dependent susceptibility:
\begin{equation}
\label{eq:kappa_tilde}
\tilde\kappa_\mathrm{NS}(s)={\partial^2M_N\over\partial s^2}=
\kappa_\mathrm{NS}\left(1+{s\over f_\pi}\right) \, ,
\end{equation} 
which vanishes at full chiral restoration, i.e., $\bar s=-f_\pi$, where $\bar s$ is the value taken by the $s$ field in the ground state. 

In addition to the scalar $s$ and vector $\omega$ fields contributing to the Hartree and Fock terms in symmetric matter, the Fock term brings a contribution from the isovector fields corresponding to the $\delta$ and $\rho$ mesons channels.

\subsection{Parameterization}

The model parameters are fixed by two parameters derived from Lattice-QCD (L-QCD) data and from hadronic phenomenology, along the lines originally proposed in Ref.~\cite{Chanfray2005}. We however need two additional constraints to determine all coupling constants, which are taken to be the empirical values of $n_\sat$ and $E_\sat$ in nuclear matter. Also notice that these parameters might be ultimately fixed or guided  by an underlying microscopic model of the QCD vacuum and of the confinement mechanism along the lines of Ref.~\cite{Chanfray2011}. The delta meson is supposed to weakly couple to nucleons and we take $g_{\delta}=1$. Note that this coupling is often neglected in relativistic approaches (see~\cite{Bouyssy_1987,Toki2010,Long2007}). As for $g_{\rho}$, we use the quark model hypothesis for which $g_{\rho} = g_{\omega}/3$, and we later on investigate the effect of replacing the quark model condition by a direct fit to the empirical value of the symmetry energy $E_\sym$, see Sec.~\ref{sec:quark model}.  The parameters considered in the present study are shown in Tabs.~\ref{tab:parameters} and \ref{tab:rhotensor}. In Tab.~\ref{tab:rhotensor}, we explore three values for the nucleon $\rho$ tensor coupling $f_\rho = g_\rho \kappa_\rho $, according to either the Vector Dominance Model (VDM)~\cite{Bhaduri1988} or to the scattering data~\cite{HOHLER1975210}. For reference we also consider the case without $\rho$ tensor coupling.

By fitting the model to the parameters in 
Tab.~\ref{tab:fitting}, i.e. using the connection to L-QCD  parameters $a_2$ and $a_4$ from the following relations (see Ref.~\cite{Massot2008,Chanfray2007,Chanfray_inprep}):
\begin{equation}
\label{eq:LQCD}
g_{s} = \frac{a_2m^2_{s}}{f_{\pi}} \quad \textrm{and} \quad C = \frac{f_{\pi}g_{s}}{M_N}\left[\frac{3}{2} + \frac{a_4m^4_{s}}{f_{\pi}g_{s}}\right]\, ,
\end{equation}
where $C$ is the dimensionless parameter related to the nucleon polarisability, i.e $C = (f^2_\pi / 2M_N)\kappa_{NS}$,
and by reproducing the Nuclear Empirical Parameters (NEP) $E_\sat$ and $n_\sat$, the following model parameters $m_{s}$, $g_{\omega}$, $g_{s}$ and $C$ can be fixed. This is shown in more details in Sec.~\ref{sec:Results}. Note that we use Bayesian statistics to propagate the experimental uncertainties into the model parameters. 

It should be remarked that, as argued in Ref.~\cite{Chanfray2011} and restated in \cite{Rahul2022}, the scalar field $s$ has no direct relation with the broad resonance $f_0$(600), but instead to the inverse of the propagator at zero momentum characterized by the scalar screening mass. As a consequence, the value of the scalar field mass does not necessarily need to match with the energy of the broad scalar meson $f_0$ of about 500~MeV (the pole mass) and can be slightly larger. This scalar screening mass is therefore running from about 600 up to about 800~MeV, namely of the order of twice the constituent quark mass in an underlying NJL model. As for $g_{s}$, a value of around 10 is predicted from the original L$\sigma$M. This value was used in Ref.~\cite{Chanfray2005,Chanfray2007,Massot2008} for instance, giving a reference value for the comparison of our new models. Note that in our case, the value of $g_{s}$ is obtained from the solution of Eqs.~\eqref{eq:LQCD}.

\begin{table}[t]
\tabcolsep=0.33cm
\def\arraystretch{1.5}
\caption{\label{tab:fitting}%
The values of the inputs to be reproduced by adjusting the remaining parameters $g_s$, $m_s$, $g_\omega$ and $C$:
the parameters $a_2$ and $a_4$ from L-QCD and the NEP $E_{\sat}$ and $n_{\sat}$. For the Lattice parameters the average and standard deviation refers to the profile of a uniform distribution, while for the NEP they correspond to a Gaussian distribution.
}
\begin{tabular}{cccc}
\hline
Parameters & Ref. & centroid & std. dev.\\
\hline
$a_2$ (GeV$^{-1}$) &  L1~\cite{LWY2003} & 1.553 & 0.136 \\
& L2~\cite{Armour_2010} & 1.1 & 0.2 \\
$a_4$ (GeV$^{-3}$) &  L1~\cite{LWY2003} & -0.509 & 0.054 \\
& L2~\cite{Armour_2010} & -0.225 & 0.05 \\
$E_{\sat}$ (MeV) & \cite{Margueron2018} & -15.8 & 0.3\\
$n_{\sat}$ (fm$^{-3}$) & \cite{Margueron2018} & 0.155 & 0.005\\ 
\hline
\label{tab:modelparameterfit}
\end{tabular}
\end{table}

Nuclear matter properties can be encoded into the nuclear empirical parameters (NEP). They are defined as the successive derivatives of the energy per particle in SM $e_{\sat}(n)$, 
\begin{equation}
P^{(k)}_{IS} = (3n_{\sat})^k \frac{\partial e_{\sat}(n)}{\partial n^k} \, ,
\end{equation}
for the iso-scalar (IS) NEPs, fixing the isospin asymmetry parameter $\delta=(n_n-n_p)/n=0$ and the density to be the saturation density, $n=n_\sat$. We note that $P^{(0)}_{IS} = E_{\sat}$ is the saturation energy, $P^{(2)}_{IS} = K_{\sat}$ is the incompressibility modulus, and $P^{(3)}_{IS} = Q_{\sat}$ is the isoscalar skewness parameter. We can also define the isovector (IV) NEPs from the symmetry energy $e_\sym(n)=e_\mathrm{NM}(n)-e_\sat(n)$, where $e_\mathrm{NM}(n)$ is the energy per particle in NM, as follows,
\begin{equation}
P^{(k)}_{IV} = (3n_{\sat})^k \frac{\partial e_{\sym}(n)}{\partial n^k} \, .
\end{equation}
We note $P^{(0)}_{IV} = E_{\sym}$ the symmetry energy at saturation, $P^{(1)}_{IV} = L_{\sym}$ the symmetry energy slope, and $P^{(2)}_{IV} = K_{\sym}$ the symmetry energy curvature. See for instance Refs.~\cite{Margueron2018,Margueron2019} for more details. 

\section{The Hartree-Fock model (model A)}
\label{sec:caseA}

In the following, the model extensions B, C and D will be based on the pure HF approach, that we present in this section. 

\subsection{Hartree-Fock equations}

From the Lagrangian expressed in Eqs.~\eqref{eq:L_kinetic} and \eqref{eq:L_meson}, and following the method proposed in~\cite{Massot2008}, we can derive the equations of motion for each meson field $\phi_\alpha$ ($\alpha=s$, $\omega$, $\rho$, $\delta$, $\pi$), giving
\begin{eqnarray}
&& -\nabla^2 s+v'(s)=
-\frac{\partial M_N}{\partial s}\,\bar\Psi\Psi\nonumber\\
&& -\nabla^2 \omega^\mu+m^{2}_{\omega}\omega^\mu+\delta_{\mu i}\partial_i(\vec{\nabla}\cdot\vec\omega)
=g_\omega\bar\Psi\gamma^\mu\Psi\nonumber\\
&& -\nabla^2 \rho_a^\mu+m^{2}_{\rho}\rho_a^\mu+\delta_{\mu i}\partial_i(\vec{\nabla}\cdot\vec\rho_a) = g_\rho\,\bar\Psi\gamma^\mu\tau_a\Psi \nonumber \\
&&\hspace{2cm} - g_\rho\frac{\kappa_\rho}{2\,M_N}\partial_j \left(\bar\Psi\sigma^{\mu j}\tau_a\Psi\right)
\nonumber\\
&& -\nabla^2 \delta_a+m^{2}_{\delta_a}\,	\delta
=g_\delta\,\bar\Psi\tau_a\Psi\nonumber\\
& & -\nabla^2 \varphi_{a\pi}\,+\,m^{2}_{\pi}\varphi_{a\pi}=
\frac{g_A}{2\,f_\pi}\,\vec{\nabla}\cdot\bar\Psi\gamma^5\vec\gamma\tau_a\Psi.
\label{eq:eom}
\end{eqnarray}
We then decompose the meson fields, see Ref.~\cite{Guichon_2006,Massot2008}, as follows,
\begin{ceqn}
\begin{equation}
\phi_\alpha = \bar{\phi}_\alpha + \Delta\phi_\alpha
\end{equation}
\end{ceqn}
where $\bar{\phi}_\alpha = \langle\phi_\alpha\rangle$ represents the  classical expectation value of the meson field $\phi_\alpha$ while $\Delta\phi_\alpha$ corresponds to its quantum fluctuation, considered here as a smaller quantity. Injecting this decomposition into the equations of motion~\eqref{eq:eom} generates two sets of equations, one describing the motion in a self consistent scalar and vector background fields (direct or Hartree terms), \begin{eqnarray}
\label{eq:classical_EOM}
&& -\nabla^2 \bar s +v'(\bar s)=-g^*_s\left\langle \bar\Psi\Psi\right\rangle\nonumber\\
&& -\nabla^2 \bar\omega^0+m^{2}_{\omega}	\bar\omega^0=g_\omega\left\langle \Psi^\dagger\Psi\right\rangle\nonumber\\
&& -\nabla^2 \bar\rho_a^0+m^{2}_{\rho}\,	\bar\rho_a^0=g_\rho\,\left\langle \Psi^\dagger\tau_a\Psi\right\rangle
-g_\rho\frac{\kappa_\rho}{2M_N}\,\partial_j \left\langle \bar\Psi\sigma^{0 j}\tau_a\Psi\right\rangle\nonumber\\
&& -\nabla^2 \bar\delta_a+m^{2}_{\delta_a}	\bar\delta
=g_\delta\left\langle \bar\Psi\tau_a\Psi\right\rangle	\, .
\end{eqnarray}
where $g^*_s=\partial M_N(\bar s) / \partial \bar s$, and the other which describes the propagation of the fluctuations of the meson fields (exchange or Fock terms), 
\begin{eqnarray}
\label{eq:Fock_EOM}
&& -\nabla^2 (\Delta s)+m^{*2}_{s}\Delta s=
-g^*_s\left(\bar\Psi\Psi-\left\langle \bar\Psi\Psi\right\rangle\right)\nonumber\\
&& -\nabla^2 (\Delta\omega^\mu)+m^{2}_{\omega}\Delta\omega^\mu=P^{\mu}_{\;\nu}
g_\omega\left(\bar\Psi\gamma^\nu\Psi-\left\langle \bar\Psi\gamma^\nu\Psi\right\rangle\right)\nonumber\\
&& -\nabla^2 (\Delta\rho_a^\mu)+m^{2}_{\rho}	\Delta\rho^\mu=	P^{\mu}_{\;\nu}
g_\rho\,\bigg(\bar\Psi\gamma^\nu\tau_a\Psi-\left\langle \bar\Psi\gamma^\nu\tau_a\Psi\right\rangle\nonumber\\
& & 
\qquad -\frac{\kappa_\rho}{2M_N}\partial_j \left(\bar\Psi\sigma^{\nu j}\tau_a\Psi\right)
+\frac{\kappa_\rho}{2M_N}\,\partial_j \left\langle\bar\Psi\sigma^{\nu j}\tau_a\Psi\right\rangle
\bigg)\nonumber\\
&& -\nabla^2 (\Delta\delta_a)+m^{2}_{\delta}\,	\Delta\delta_a
=g_\delta\left(\bar\Psi\tau_a\Psi-\left\langle \bar\Psi\tau_a\Psi\right\rangle\right)\nonumber\\
&& -\nabla^2 (\Delta\varphi_{a\pi})+m^{2}_{\pi}\,\Delta\varphi_{a\pi}
=\frac{g_A}{2\,f_\pi}\vec{\nabla}\cdot\bar\Psi\gamma^5\vec\gamma\tau_a\Psi
\end{eqnarray}
with $m^{*2}_{s}=v''(\bar s)\,+\,\tilde{\kappa}_\mathrm{NS}\,\left\langle \bar\Psi\Psi\right\rangle$ and the operator $P$ defined as,
\begin{equation}
P^{0}_{\;0}=1,\, P^{0}_{\;i}=0,\, P^{i}_{\;0}=0,\, P^{i}_{\;j}\equiv P^{i}_{\;j}=\delta_{ij}\,-\,\frac{\partial_i\partial_j}{m^{2}_{\rho}}.
\end{equation}

\begin{table*}[t]
\tabcolsep=0.3cm
\def\arraystretch{2.0}
\caption{\label{tab:coefficients}Functions $A_\alpha$, $B_\alpha$, $C_\alpha$ and $D_\alpha$ used in the expressions for the exchange self-energies.}
\begin{tabular}{c c c c c} 
\hline
$\alpha$  & $A_\alpha$ & $B_\alpha$ & $C_\alpha$ & $D_\alpha$ \\ [0.5ex] 
\hline
$s$ & $g^{*2}_{s}\theta_{s}$ & $g^{*2}_{s}\theta_{s}$ & $-2g^{*2}_{s}\phi_{s}$ & - \\
$\delta$ & $g^2_{\delta}\theta_{\delta}$ & $g^2_{\delta}\theta_{\delta}$ & $-2g^2_{\delta}\phi_{\delta}$ & - \\ 
$\omega$ & $2g^2_{\omega}\theta_{\omega}$ & $-4g^2_{\omega}\theta_{\omega}$ & $-4g^2_{\omega}\phi_{\omega}$ & - \\
$\rho^V$ & $2g^2_{\rho}\theta_{\rho}$ & $-4g^2_{\rho}\theta_{\rho}$ & $-4g^2_{\rho}\theta_{\rho}$ & - \\ 
$\rho^T$ & $-\left(\frac{f_\rho}{2M}\right)^2m^2_{\rho}\theta_{\rho}$ & $-3\left(\frac{f_\rho}{2M}\right)^2m^2_{\rho}\theta_{\rho}$ & $4\left(\frac{f_\rho}{2M}\right)^2m^2_{\rho}[(\mathrm{p}^2+\mathrm{p}^{\prime 2}-m^2_{\rho}/2)\phi_{\rho} - \mathrm{p}\mathrm{p}^\prime\theta_{\rho}]$ & - \\
$\rho^{VT}$ & - & - & - & 12$\left(\frac{g_\rho f_\rho}{2M}\right)(\mathrm{p}\theta_{\rho} - 2\mathrm{p}^\prime\phi_{\rho})$ \\ 
$\pi$ & -$ \left(\frac{g_A}{2f_\pi}\right)^2m^2_\pi \theta_{\pi}$  &  $-  \left(\frac{g_A}{2f_\pi}\right)^2m^2_\pi \theta_{\pi} $ & $2 \left(\frac{g_A}{2f_\pi}\right)^2[(\mathrm{p}^2+\mathrm{p}^{\prime 2})\phi_{\pi}-\mathrm{p}\mathrm{p}^{\prime}\theta_{\pi}] $ & - \\
\hline
\end{tabular}
\end{table*}

The Hartree-Fock Hamiltonian is determined from the Lagrangians~\eqref{eq:L_kinetic} and \eqref{eq:L_mesons}, and the equations of motion decomposed into the direct and the exchange component of the fields. It can be written in two terms,
\begin{equation}
H\equiv H^{K+D}\,+\,H^{E},
\end{equation}
where the first term contains the kinetic (K) and direct or Hartree (D) contributions,
\begin{eqnarray}
H^{K+D} &=& \int \!\!d{\bf r}\bigg[\bar\Psi\bigg(-i\vec{\gamma}\cdot\vec{\nabla}+M_N(\bar s)+g_\omega
\bar\omega^0\gamma_0
\nonumber\\
&&\hspace{-0.5cm}+g_\rho\bar\rho_3^0\gamma_0\tau_3
+g_\rho\frac{\kappa_\rho}{2M_N}\partial_j \bar\rho^0_3\sigma^{0 j}\tau_3+g_\delta
\bar\delta_3\tau_3\bigg)\Psi \nonumber \\
&&\hspace{-0.5cm}+v(\bar s)+\frac{1}{2}\,
\left(\vec{\nabla}\bar s\right)^2 
+\frac{1}{2}m^2_\delta\bar\delta_3^2
+\frac{1}{2}\left(\vec{\nabla}\bar\delta_3\right)^2 \nonumber\\
&&\hspace{-0.5cm}-\frac{1}{2}m^2_\omega (\bar\omega^0)^2-
\frac{1}{2}\left(\vec{\nabla}\bar\omega^0\right)^2 \nonumber \\
&&\hspace{-0.5cm}-\frac{1}{2}m^2_\rho(\bar\rho_3^0)^2 -\frac{1}{2}\left(\vec{\nabla}\bar\rho^0_3\right)^2 \bigg] \, ,
\end{eqnarray}
and the second term $H^E$ corresponds to the exchange mediated by the propagation of the meson fields' fluctuations. It has the form
\begin{eqnarray}
H^E &=&	\int\!\! d{\bf r}\frac{1}{2}\,\bigg[g^*_s\Delta s\Delta\left(\bar\Psi\Psi\right)
+g_\omega\Delta \omega_\mu\left(\bar\Psi\gamma^\mu\Psi\right)\nonumber\\
&&\hspace{-0.5cm}+g_\rho\Delta\rho^{a}_\mu\bigg(\Delta(\bar\Psi\gamma^\mu\tau_a\Psi)-
\frac{\kappa_\rho}{2M_N}\partial_j \left[\Delta\left(\bar\Psi\sigma^{\mu j}\tau_a\Psi\right)\right]\bigg)\nonumber\\
&&\hspace{-0.5cm}+g_\delta\Delta\delta_a\Delta\left(\bar\Psi\tau_a\Psi\right)
+\frac{g_A}{2f_\pi}\Delta\varphi_{a\pi}\vec{\nabla}\cdot\bar\Psi\gamma^5\vec\gamma\tau_a\Psi\bigg]
\end{eqnarray}
where we use the following notation,
$$\Delta(\bar\Psi\Gamma\Psi)=\bar\Psi\Gamma\Psi-\left\langle \bar\Psi\Gamma\Psi\right\rangle\,.$$

We introduce the (static) propagators $D_\alpha({\bf r}-{\bf r}')$ in coordinate space for the fluctuating fields, defined as,
\begin{eqnarray}
\left(-\nabla^{2}_{\bf r}+m^{*2}_{s}({\bf r})\right)D_s({\bf r}-{\bf r}')&=&\delta^{(3)}({\bf r}-{\bf r}'),\nonumber\\
\left(-\nabla^{2}_{\bf r}+m^{2}_{\alpha}\right)D_\alpha({\bf r}-{\bf r}') &=&	\delta^{(3)}({\bf r}-{\bf r}')\, ,
\end{eqnarray}
with $\alpha=\omega$, $\rho$, $\delta$, and $\pi$.
The exchange term $H^{E}$ can then be written as,
\begin{eqnarray}
\label{eq:exchange_hamiltonian}
H^E&=& \frac{1}{2}\int\!\! d{\bf r} d{\bf r}' \bigg[
-g^*_s({\bf r}) g^*_s({\bf r}') \Delta\left(\bar\Psi\Psi\right)({\bf r}) \nonumber\\
&& \hspace{2cm}\times D_s({\bf r}-{\bf r}')
\Delta\left(\bar\Psi\Psi\right)({\bf r}')\nonumber\\
&&+g^{2}_\omega \Delta\left(\bar\Psi\gamma^\mu\Psi\right)({\bf r})
D_{\omega,\mu\nu}({\bf r}-{\bf r}')
\Delta\left(\bar\Psi\gamma^\nu\Psi\right)({\bf r}')\nonumber\\
&&+g^{2}_\rho \Delta\left(\bar\Psi\gamma^\mu\tau_a\Psi\right)({\bf r})
D_{\rho,\mu\nu}({\bf r}-{\bf r}')
\Delta\left(\bar\Psi\gamma^\nu\tau_a\Psi\right)({\bf r}')\nonumber\\
&&+2g^{2}_\rho\frac{\kappa_\rho}{2\,M_N} \Delta\left(\bar\Psi\sigma^{\mu j}\tau_a\Psi\right)({\bf r}) \nonumber\\
&& \hspace{2cm}\times \partial_j D_{\rho,\mu\nu}({\bf r}-{\bf r}') \Delta\left(\bar\Psi\gamma^\nu\tau_a\Psi\right)({\bf r}')\nonumber\\
&& +g^{2}_\rho\left(\frac{\kappa_\rho}{2 M_N}\right)^2 \Delta\left(\bar\Psi\sigma^{\mu i}\tau_a\Psi\right)({\bf r}) \nonumber\\
&& \hspace{2cm}\times \partial_i\partial'_j D_{\rho,\mu\nu}({\bf r}-{\bf r}')
\Delta\left(\bar\Psi\sigma^{\nu j}\tau_a\Psi\right)({\bf r}')\nonumber\\
&&-g^2_\delta \Delta\left(\bar\Psi\tau_a\Psi\right)({\bf r}) D_\delta({\bf r}-{\bf r}')
\Delta\left(\bar\Psi\tau_a\Psi\right)({\bf r}')\nonumber\\
&& +\left(\frac{g_A}{2\,f_\pi}\right)^2\left(\bar\Psi\gamma^5\gamma^i\tau_a\Psi\right)({\bf r})
\partial_i\partial'_j D_\pi({\bf r}-{\bf r}') \nonumber\\
&& \hspace{2cm}\times \left(\bar\Psi\gamma^5\gamma^j\tau_a\Psi\right)({\bf r}')\bigg]
\end{eqnarray}
where we have introduced the tensor propagator $D_{\omega,\mu\nu}({\bf r}-{\bf r}')$ whose non vanishing components are~:
$D_{\omega,00}({\bf r}-{\bf r}')=D_{\omega}({\bf r}-{\bf r}')$ and
$D_{\omega,ij}({\bf r}-{\bf r}')=(\delta_{ij}-\partial_{i}\partial_{j}/m^2_\omega)D_{\omega}({\bf r}-{\bf r}')$ and a similar one for the $\rho$ meson. 
 
In the Hartree-Fock (HF) approximation, the ground state is represented by a Slater determinant composed of single-particle Dirac wave functions $\varphi_a^N(\vec{r})~\chi_N$, where $\chi_N$ is the spin wave function and $N=n$ (neutrons) or $p$ (protons). We can define the densities appearing in the sources of the classical equations of motions as
\begin{eqnarray}
\left\langle \bar{\Psi}\Psi\right\rangle &=& \sum_{a<F}\,\bar{\varphi}_a^{p}\varphi^{p}_{a}\,+\,
\bar{\varphi}_a^{n}\varphi^{n}_{a}\equiv n_{sp}\,+\,n_{sn}=n_s\nonumber\\
\left\langle \Psi^\dagger\Psi\right\rangle &=& \sum_{a<F}\,\varphi_a^{p\dagger}\varphi^{p}_{a}\,+\,
\varphi_a^{n\dagger}\varphi^{n}_{a}\equiv n_{p}\,+\,n_{n}=n\nonumber\\
\left\langle \bar{\Psi}\tau_3\Psi\right\rangle &=& \sum_{a<F}\,\bar{\varphi}_a^{p}\varphi^{p}_{a}\,-\,
\bar{\varphi}_a^{n}\varphi^{n}_{a}\equiv n_{sp}\,-\,n_{sn}=n_s^{(3)}\nonumber\\
\left\langle \Psi^\dagger\tau_3\Psi\right\rangle &=& \sum_{a<F}\,\varphi_a^{p\dagger}\varphi^{p}_{a}\,-\,
\varphi_a^{n\dagger}\varphi^{n}_{a}\equiv n_{p}\,-\,n_{n}=n^{(3)}\nonumber\\
\left\langle \bar{\Psi}\sigma^{0 j}\tau_3\Psi\right\rangle &=& \sum_{a<F}\,\bar{\varphi}_a^{p}\sigma^{0 j}\varphi^{p}_{a}\,-\,
\bar{\varphi}_a^{n}\sigma^{0 j}\varphi^{n}_{a}.
\end{eqnarray}

The single-particle orbitals are obtained by minimizing the HF energy with respect to the $\bar{\varphi}_a^{p,n}(\vec{r})$, with the constraint that the single-particle wave functions are normalized, which could be expressed in the following way by introducing Lagrange multipliers $\epsilon_a^N$,
\begin{ceqn}
\begin{equation}
    \frac{\delta}{\delta\bar{\varphi}_a^N(\mathbf{r})} \left(E-\sum_{N,a}\epsilon_a^N\int{d\mathbf{r}^\prime}\varphi_a^{N\dagger}(\mathbf{r}^\prime)\varphi_a^N(\mathbf{r}^\prime)\right) = 0 \, .
\end{equation}
\end{ceqn}
The Lagrange parameters $\epsilon_a^N$ associated to the normalisation of the wave-functions stand for the single-particle energies. When studying infinite nuclear matter, the single-particle orbitals are plane waves, labeled by momentum and spin indices $p=(\mathbf{p},s)$ for each isospin state $N$:
\begin{ceqn}
\begin{equation}
\varphi^N_p(\mathbf{r}) = \frac{1}{\sqrt{V}}u(\mathbf{p},s)e^{i\mathbf{p}\cdot\mathbf{r}}\chi_N \, ,
\end{equation}
\end{ceqn}
where $V$ is the the volume of a unit cell.

The Dirac spinors $u(\mathbf{p},s)$ minimizing the energy are defined as the solution of the following Dirac equation:
\begin{ceqn}
\begin{equation}
\label{eq:Dirac}
\left[ \bm{\gamma}.\mathbf{p}^* + M^*_D \right] u(\mathbf{p},s) = \gamma_0E^*u(\mathbf{p},s) 
\end{equation}
\end{ceqn}
where the starred quantities are the effective momentum $\mathbf{p}^*$, the Dirac scalar mass $M^*_D$, and the effective energy $E^*$, given by,
\begin{ceqn}
\begin{eqnarray}
\mathbf{p}^* &=& \mathbf{p} + \mathbf{\hat{p}}\Sigma_V(\mathrm{p})\ , \nonumber \\
M^*_D(\mathrm{p}) &=& M_N + \Sigma_S(\mathrm{p})\ , \\
E^*(\mathbf{p}) &=& \epsilon^N_a - \Sigma_0(\mathrm{p})=\sqrt{M^{*}_D(\mathrm{p})^2 +\mathbf{p}^{*2}}\, , \nonumber
\end{eqnarray}
\end{ceqn}
where $\Sigma_S(\mathrm{p})$, $\Sigma_0(\mathrm{p})$, and $\Sigma_V(\mathrm{p})$ are the self-energies of scalar, time and vector nature, and they receive contributions from all the mesons we consider in our study, namely $s$, $\delta$, $\omega$, $\rho$ and $\pi$, as well as from the direct and exchange contributions. These quantities can be different for protons and neutrons in isospin asymmetric matter. Note that $\mathrm{p}=\vert\mathbf{p}\vert$.

The natural definition of the in-medium kinetic energy-density is,
\begin{equation}
\label{eq:KE_original}
\epsilon^K_\mathrm{nat} = \sum\limits_{N=p,n} \frac{1}{\pi ^2} \int_{0}^{k_{F_N}}\!\!d\mathrm{p} \mathrm{p}^2\left(\mathrm{p} \hat{P}_N + M_N(\bar s) \hat{M}_N \right)
\end{equation}
where the quantities $\hat{M}_N$ and $\hat{P}_N$ are defined as,
\begin{ceqn}
\begin{equation} \label{eq:hatted_quantities}
\hat{M}_N(\mathrm{p})\equiv \frac{M^*_{DN}(\mathrm{p})}{E^*_N(\mathrm{p})}
\qquad
\hat{P}_N(\mathbf{p}) \equiv \frac{\mathbf{p}^*_N}{E^*_N(\mathrm{p})}, \end{equation}
\end{ceqn}
and the scalar meson Hartree energy is simply
\begin{equation}
\epsilon^{H,s}_\mathrm{nat} = v(\bar s) \, .
\end{equation}
In practice, the in-medium kinetic energy \eqref{eq:KE_original} gets unexpected negative values since $M_N(\bar{s})$ is density dependent and decreases as the density increases. We therefore adopt the following definition of the kinetic energy
\begin{equation}
\epsilon^K = \sum\limits_{N=p,n} \frac{1}{\pi ^2} \int_{0}^{k_{F_N}}\!\!\mathrm{p}^2d\mathrm{p} \left(\mathrm{p} \hat{P}_N + M_N \hat{M}_N \right) \, ,
\end{equation}
while the scalar meson Hartree contribution to the energy becomes
\begin{equation}
\epsilon^{H,s} = \left(M_N(\bar s) - M_N\right) n_s+ v(\bar s) \, ,
\end{equation}
such that sum of the kinetic and scalar Hartree terms is unmodified: $\epsilon^K_\mathrm{nat}+\epsilon^{H,s}_\mathrm{nat}=\epsilon^K+\epsilon^{H,s}$.

The general expression for the Hartree and Fock energies per particle, respectively $\epsilon^H$ and $\epsilon^F$, are given by:
\begin{subequations}
\begin{eqnarray}
\epsilon^H &=& \left(M_N(\bar s) - M_N\right) n_s + v(\bar s) 
- \frac{1}{2} \left(\frac{g_{\delta}}{m_{\delta}} \right)^2 (n_{sn} - n_{sp})^2
\nonumber \\
&& + \frac{1}{2} \left(\frac{g_{\omega}}{m_{\omega}} \right)^2 n^2  + \frac{1}{2} \left(\frac{g_{\rho}}{m_{\rho}} \right)^2 (n_n - n_p)^2 \, , \\ 
\epsilon^F &=& \int\!\!\frac{d\mathbf{p}}{(2\pi)^3}\!\!\sum\limits_{N=p,n} \big(\Sigma^E_{0,N}
+ \hat{M}_N(\mathrm{p})\Sigma^E_{S,N} \nonumber \\
&& + \hat{P}_N(\mathrm{p})\Sigma^E_{V,N}\big) \, , 
\label{eq:energy_per_particle}
\end{eqnarray}
\end{subequations}
where the various $\Sigma^E$ are the exchange part of the self-energies, and the densities in uniform matter become
\begin{eqnarray}
n_n &=& \int\frac{2\,d{\bf p}}{(2\pi)^3}\,f_n(\mathrm{p}), \,
n_{sn} = \int\frac{2\,d{\bf p}}{(2\pi)^3}\,\hat{M}_n(\mathrm{p})\,f_n(\mathrm{p})\, , \\
n_p &=&\int\frac{2\,d{\bf p}}{(2\pi)^3}\,f_p(\mathrm{p}), \,
n_{sp} = \int\frac{2\,d{\bf p}}{(2\pi)^3}\,\hat{M}_p(\mathrm{p})\,f_p(\mathrm{p})\, ,
\end{eqnarray}
where $f_N(\mathrm{p}) = \theta(\mathrm{p}_{F_N} - \mathrm{p})$ is the occupation number for the nucleon $N$ characterized by the Fermi momentum $p_{F_N}$. 

The direct (Hartree) contribution to the self-energies can be found in \ref{app:direct}. Note that these direct contributions are identical for all the models we study in this paper. In the following we concentrate on the exchange contributions to the self-energies. 

For the scalar $s$ field, the expressions for these exchange self-energies are given by
\begin{subequations}
\begin{eqnarray}
\Sigma^{E,s}_{S,N} &=& \frac{1}{4}g_{s}^{*^2} \!\int \!\! \frac{d\mathbf{p}'}{(2\pi)^3}D_s(\mathrm{q}) 2\hat{M}_N(\mathrm{p}^\prime) + \Sigma_S^{rg} \, ,\label{eq:selfS}\\
\Sigma^{E,s}_{0,N} &=& \frac{1}{4}g_{s}^{*^2} \!\int \!\!\frac{d\mathbf{p}'}{(2\pi)^3}D_s(\mathrm{q}) 2f_N(\mathrm{p}^\prime)\, , \label{eq:self0} \\
\Sigma^{E,s}_{V,N} &=& -\frac{1}{4}g_{s}^{*^2} \!\int\!\!\frac{d\mathbf{p}'}{(2\pi)^3} D_s(\mathrm{q}) 2\tilde{\bf p} \cdot \hat{P}_N(\mathbf{p}^\prime)\, , \label{eq:selfV} 
\end{eqnarray}
\end{subequations}
where we introduce the unit momentum vector $\tilde{\bf p} = \frac{{\bf p}}{ \mathrm{p}}$, and the exchange momentum $\mathbf{q}=\mathbf{p}-\mathbf{p}^\prime$ with $\mathrm{q}=\vert\mathbf{q}\vert$. The interaction element $D_s(\mathrm{q})=[\mathrm{q}^2+m_s^{*2}]^{-1}$ is also the in-medium $s$ propagator. The term $\Sigma_S^{rg}$ appearing in the scalar self energy is the rearrangement term. It is due to the dependence of both the coupling constant and mass on the scalar field, as discussed in details in \cite{Massot2008}, and we show the expression in \eqref{eq:rg1} and \eqref{eq:rg2}. The exchange self-energies for all mesons are given in \ref{app:selfA}.

\begin{table*}[t]
\tabcolsep=0.075cm
\def\arraystretch{2.0}
\caption{\label{tab:results_A} The parameters for the model A fitted to the mean value $L1$ of the L-QCD parameters $a_2$ and $a_4$ and considering the different values of the $\rho^T$ coupling $\kappa_{\rho}$: NRT, WRT, SRT. We show the nuclear empirical parameters $K_\sat$ and $E_\sym$, as well as the scalar Dirac mass $M^*_{D}/M_N$. We also show the kinetic energy $E_K$ and the Hartree and Fock energy contributions of the scalar field $s$ and of the $\omega$ and $\rho$ meson as well as the
Fock contribution from the $\pi$ meson at in symmetric matter. Results at the Hartree approximation, named RH$_\mathrm{CC}$, are shown in the first row as a reference. The small contribution of the delta meson is not shown.}
\begin{tabular}{lcccc|cccc|cccccccc}
\hline
\multicolumn{5}{c|}{Parameters} & \multicolumn{4}{c}{NEP}  & \multicolumn{8}{|c}{Meson contribution to the binding energy} \\
\hline
model & $m_{s}$ & $g_{s}$ & $g_{\omega}$ & $C$ & $K_{\sat}$ & $E_{\sym}$ & $M_{D}^*/M_N$ & $E_\mathrm{K}$ & $E^{s}_\mathrm{H}$  & $E^{s}_\mathrm{F}$  & $E^{\omega}_\mathrm{H}$ & $E^{\omega}_\mathrm{F}$ & $E^{\pi}_\mathrm{F}$ & $E^{\rho^V}_\mathrm{F}$ & $E^{\rho^T}_\mathrm{F}$ & $E^{\rho^{VT}}_\mathrm{F}$ \\
 & MeV & & & & MeV & MeV & & MeV & MeV & MeV & MeV & MeV & MeV & MeV & MeV & MeV \\
\hline
RH$_\mathrm{CC}(L1)$~\cite{Rahul2022} & 825 & 11.092 & 6.341 & 1.41 & 264 & 18.05 & 0.86 & 24.8 & -79.7 & 0.0 & 39.1 & 0.0 & 0.0 & 0.0 & 0.0 & 0.0 \\
RHF$_\mathrm{CC}(A,L1,\mathrm{NRT})$ & 982 & 15.743 & 7.936 & 1.86 & 326 & 23.7 & 0.73 & 19.1 & -109.2 & 14.8 & 61.2 & -11.6 & 13.4 & -3.9 & 0.0 & 0.0 \\
RHF$_\mathrm{CC}(A,L1,\mathrm{WRT})$ & 941 & 14.448 & 6.810 & 1.74 & 315 & 21.3 & 0.78 & 19.5 & -99.6 & 13.1 & 45.0 & -8.6 & 13.5 & -2.9 & 1.8 & 2.0 \\
RHF$_\mathrm{CC}(A,L1,\mathrm{SRT})$ & 911 & 13.527 & 5.839 & 1.66 & 306 & 19.7 & 0.81 & 20.0 & -93.0 & 11.9 & 33.1 & -6.4 & 13.5 & -2.1 & 4.3 & 2.5 \\
\hline
\end{tabular}
\end{table*}

The angular integration in Eqs.~\eqref{eq:selfS}, \eqref{eq:self0} and \eqref{eq:selfV} can be performed using
\begin{subequations}
\begin{align}
\int \frac{d\Omega}{\mathbf{q}^2 + m^2_\alpha} &= \frac{2\pi}{2\mathrm{p} \mathrm{p}^\prime }\theta_\alpha(\mathrm{p}, \mathrm{p}^\prime) \, , \\
\int \frac{\cos{\theta}d\Omega}{\mathbf{q}^2 + m^2_\alpha} &=
\frac{2\pi}{\mathrm{p} \mathrm{p}^\prime }\phi_\alpha(\mathrm{p}, \mathrm{p}^\prime) \, , \\
\int \frac{\cos^2{\theta}d\Omega}{\mathbf{q}^2 + m^2_\alpha} &=
2\pi\frac{\mathrm{p}^2+\mathrm{p}^{\prime 2}+m^2_\alpha}{2\mathrm{p}^2 \mathrm{p}^{\prime 2} }\phi_\alpha(\mathrm{p}, \mathrm{p}^\prime) \, ,
\end{align}
\end{subequations}
where $\mathbf{p} \cdot \mathbf{p}^\prime=\mathrm{p} \mathrm{p}^\prime \cos\theta$ and
the functions $\theta_\alpha$ and $\phi_\alpha$ are defined as,
\begin{subequations}
\begin{align}
\theta_\alpha(\mathrm{p}, \mathrm{p}^\prime) &= \ln \left (\frac{(\mathrm{p} + \mathrm{p}^\prime)^2 + m^2_\alpha}{(\mathrm{p} - \mathrm{p}^\prime)^2 + m^2_\alpha} \right )\, ,  \label{eq:theta} \\
\phi_\alpha(\mathrm{p}, \mathrm{p}^\prime) &= \frac{1}{4\mathrm{p}\mathrm{p}^\prime}(\mathrm{p}^2 + \mathrm{p}^{\prime 2} + m^2_\alpha)\theta_\alpha(\mathrm{p}, \mathrm{p}^\prime) - 1 \, . \label{eq:phi}
\end{align}
\end{subequations}
We obtain for the exchange self-energies:
\begin{eqnarray}
\Sigma^{E,\alpha}_S(\mathrm{p},\tau) &=& \langle \hat{M}(\mathrm{p}^\prime)B_\alpha(\mathrm{p}, \mathrm{p}^\prime)+\frac{1}{2}\hat{P}(\mathbf{p}^\prime)D_\alpha(\mathrm{p}^\prime, \mathrm{p})\rangle \nonumber \\
&&+ \Sigma_S^{E,c,\alpha}, \label{eq:sigma_S} \\
\Sigma^{E,\alpha}_0(\mathrm{p},\tau) &=& \langle A_\alpha(\mathrm{p}, \mathrm{p}^\prime)\rangle + \Sigma_0^{E,c,\alpha}, \label{eq:sigma_0} \\
\Sigma^{E,\alpha}_V(\mathrm{p},\tau) &=& \langle \hat{P}(\mathbf{p}^\prime)C_\alpha(\mathrm{p}, \mathrm{p}^\prime)+\frac{1}{2}\hat{M}(\mathbf{p}^\prime)D_\alpha(\mathrm{p}, \mathrm{p}^\prime)\rangle \nonumber \\
&&+ \Sigma_V^{E,c,\alpha}, \label{eq:sigma_V} 
\end{eqnarray}
using the following notation for any function $X_\alpha (\mathrm{p}, \mathrm{p}^\prime)$,
\begin{equation}
\langle X_\alpha(\mathrm{p}, \mathrm{p}^\prime) \rangle = \frac{1}{(4\pi)^2 \mathrm{p}}\sum\limits_{\alpha,\tau'}\tau_\alpha^2
\int_{0}^{p_{F_N}}\!\!\mathrm{p}^\prime d\mathrm{p}^\prime X_\alpha(\mathrm{p}, \mathrm{p}^\prime) \, ,
\end{equation}
with $\tau_\alpha$ the $N$ isospin factor at the meson $\alpha$ vertex and the expressions for $A_\alpha$, $B_\alpha$, $C_\alpha$ and $D_\alpha$ are given in Tab.~\ref{tab:coefficients}.

The contact exchange self-energies $\Sigma^{E,\mathrm{c}}$ correspond to the contact part of the interaction for the $\pi$ and $\rho^T$ interactions, which is isolated by rewriting the interaction element in these terms as follows:
\begin{equation}
\label{eq:V_A}
 V_{A,\alpha}(\mathrm{q})=\frac{\mathrm{q}^2}{\mathrm{q}^2 + m^2_\alpha} 
= 1-\frac{m^2_\alpha}{\mathrm{q}^2 + m^2_\alpha} \quad (\alpha = \pi,\rho^T)
\end{equation}
and thus
\begin{subequations}
\begin{eqnarray}
\Sigma_S^{E,c} &=& \left(\frac{g_A}{2f_\pi}\right)^2 \langle 4\mathrm{p}\mathrm{p}^\prime\hat{M}(\mathrm{p}^\prime) \rangle +
\left(\frac{f_{\rho}}{2M}\right)^2 \langle 12\mathrm{p}\mathrm{p}^\prime \hat{M}(\mathrm{p}^\prime) \rangle \, , \nonumber \\ \label{eq:selfSc}\\
\Sigma_0^{E,c} &=& \left(\frac{g_A}{2f_\pi}\right)^2 \langle 4\mathrm{p}\mathrm{p}^\prime \rangle + 
\left(\frac{f_{\rho}}{2M}\right)^2 \langle 4\mathrm{p}\mathrm{p}^\prime \rangle \, , \label{eq:self0c}\\
\Sigma_V^{E,c} &=& 0 \label{eq:selfVc} \, . 
\end{eqnarray}
\end{subequations}

\subsection{Discussion of the results}

We compare in Tab.~\ref{tab:results_A} the Hartree-Fock calculations for the different choices of the $\rho^T$ coupling (NRT, WRT, SRT) (see Tab.~\ref{tab:rhotensor}) to the pure Hartree one, which is named RH$_\mathrm{CC}$($L1$). Note that the latter is identical to the RMF-CC models analysed in the paper~\cite{Rahul2022}. Since it has no Fock term, this model is not impacted by the choice of the $\rho^T$ coupling neither by the way the SRC are treated. It is therefore independent of the case A to D that we explore in the next sections. In the fitting protocol of all these models, we employ the centroid value for the distribution of the parameters $a_2$ and $a_4$ as defined by $L1$, see Tab.~\ref{tab:fitting}. The main difference between the RH$_\mathrm{CC}$ and the RHF$_\mathrm{CC}$ models lies in the $s$ channel: $m_s$, $g_s$ as well as $g_s/m_s$ are systematically lower in the RH$_\mathrm{CC}$ model compared to the RHF$_\mathrm{CC}$ models for the three choices of the $\rho^T$ coupling constants. The lower value for $C$ (and therefore for the incompressibility $K_\sat$) for the RH$_\mathrm{CC}$ model compared to the RHF$_\mathrm{CC}$ models is a consequence of the values taken by $m_s$ and $g_s$, see Eqs.~\eqref{eq:LQCD}. Note that $g_\omega$ is not fixed by Eqs.~\eqref{eq:LQCD} but only by the constraint to reproduce saturation energy and density.

Concerning the NEP, the incompressibility modulus $K_\sat$ are larger in the RHF$_\mathrm{CC}$ models in comparison to the the value obtained in the RH$_\mathrm{CC}$ case, making it larger than the expected value of $230\pm 20$~MeV~\cite{Margueron2018}. The symmetry energy $E_\sym$ is also larger for  RHF$_\mathrm{CC}$ compared to RH$_\mathrm{CC}$ case, but since the value of the RH$_\mathrm{CC}$ case is lower than the expected one, i.e. $32\pm 2$~MeV~\cite{Margueron2018}, this effect leads to an improvement of the symmetry energy. Note also the role of the $\rho^T$ coupling: increasing the $\rho^T$ coupling leads to a decrease of the symmetry energy. The scalar Dirac mass $M^*_{D}$ is also larger for RH$_\mathrm{CC}$ compared to RHF$_\mathrm{CC}$. As a consequence, the kinetic energy is also larger for RH$_\mathrm{CC}$ compared to RHF$_\mathrm{CC}$: The kinetic energy increases as $M^*_{D}$ increases for the three models explored with RHF$_\mathrm{CC}$.

We also detail in Tab.~\ref{tab:results_A} the contribution of the different mesons to the binding energy in symmetric matter (except the $\delta$ meson which gives a small contribution). It is reminded that all the models are calibrated to reproduce the same value of the binding energy $E_\sat$ at saturation density $n_\sat$, see Tab.~\ref{tab:fitting}. The Hartree energy of both the $s$ field and $\omega$ mesons are larger in absolute value for RHF$_\mathrm{CC}$ models compared to the RH$_\mathrm{CC}$ one. The Fock contributions of these mesons are opposite in sign to the Hartree terms, and represent about 12 to 18\% of the Hartree contribution. These Fock terms contribute to about half of the difference between RHF$_\mathrm{CC}$ and RH$_\mathrm{CC}$ models. The other half is given dominantly by the (repulsive) pion. The contribution of the $\rho$ meson remains small, about 1 to 5\% in absolute value of the $s$ Hartree term: it is attractive (repulsive) and contributes to about -4 (+4)~MeV in the absence (presence) of $\rho^V$.

This analysis of the pure HF model shows that the effect of the Fock term is to moderately increase the incompressibility modulus (by 15 to 20\%), but at the same time it also contributes to increase to symmetry energy (by 10 to 30\%) and therefore it makes it closer to its empirical value. There is an effect of the considered $\rho^T$ coupling: an increase of this coupling leads to a decrease in $K_\sat$ and $E_\sym$. In general, we obtain an improvement in the reproduction of the experimental NEPs by including the Fock term to the total energy. 

\begin{figure}[t]
\centering
\includegraphics[width=0.5\textwidth]{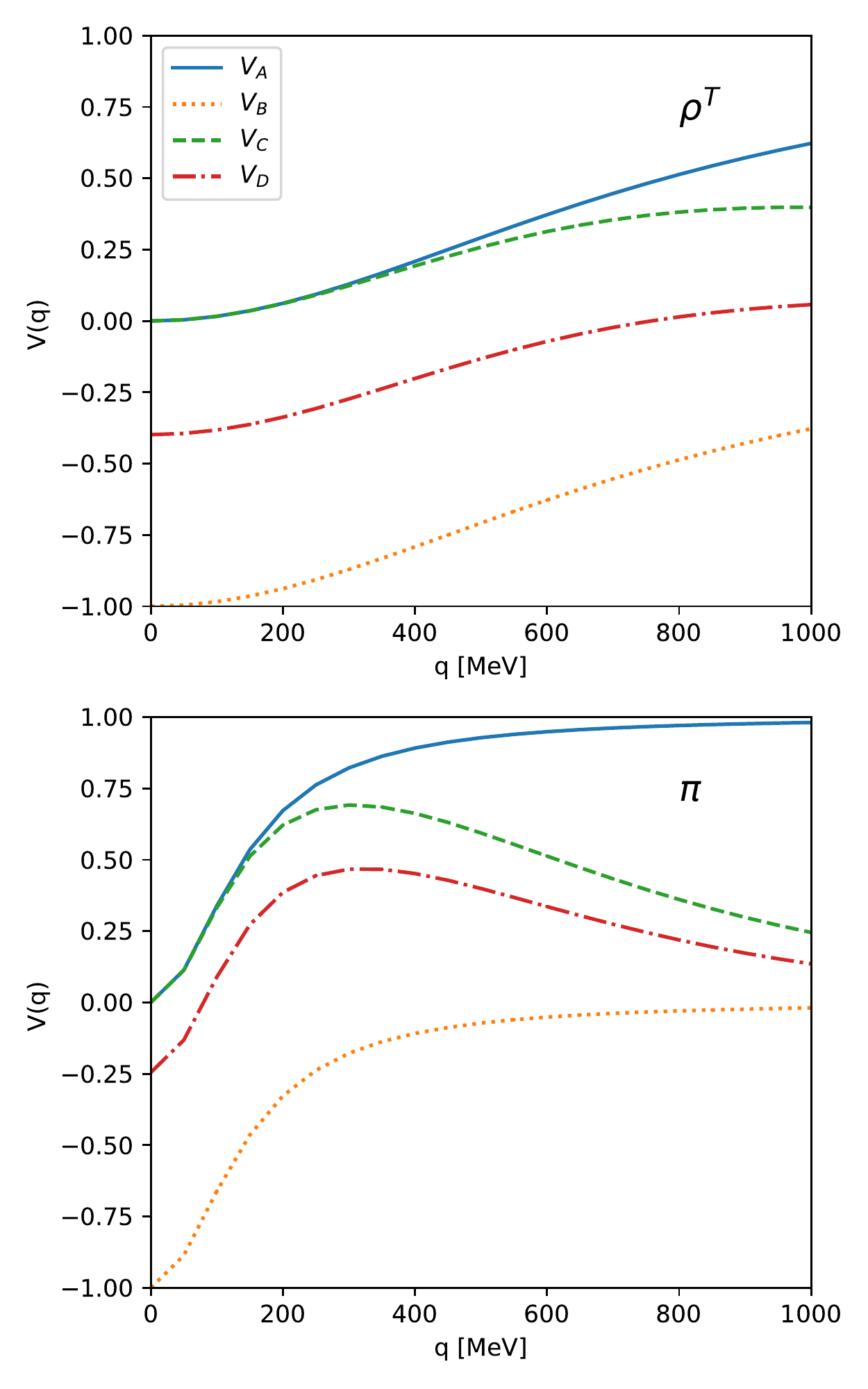}
\caption{The $\pi$ and $\rho^T$ meson exchange interaction elements $V_\alpha(\mathrm{q})$ for different cases: $V_{A,\alpha}(\mathrm{q})$ is the pure HF in-medium interaction (solid line), $V_{B,\alpha}(\mathrm{q})$ the interaction element obtained from the Orsay prescription (dotted line), $V_{C,\alpha}(\mathrm{q})$ with the form factor correction (dashed line) and $V_{D,\alpha}(\mathrm{q})$ the form factor and Jastrow SRC (dashed-dotted line). The $\rho^T$ ($\pi$) interaction is shown on the top (bottom) panel. The FF is a monopole type factor where $\Lambda_{\rho^T}$ = 2000 MeV and $\Lambda_{\pi}$ = 1000 MeV and we took $\mathrm{q}_c$ = 1000 MeV for the JSRC.}
\label{fig:V(q)}
\end{figure}

\begin{table}[t]
\tabcolsep=0.7cm
\def\arraystretch{1.3}
\caption{Dimensionless spin-isospin Landau-Midgal parameter $g^\prime$ for the models A to D and for the set $L1$ obtained from L-QCD.}
\label{tab:lmg}
\begin{tabular}{c|ccc}
\hline
Model & NRT & WRT & SRT \\
\hline
A  & $0.00$ & $0.00$ & $0.00$ \\
B  & $0.33$ & $0.78$ & $2.10$ \\
C  & $0.00$ & $0.00$ & $0.00$ \\
D & $0.08$ & $0.23$ & $0.48$ \\
\hline
\end{tabular}
\end{table}

It is also interesting to analyze the prediction of the pure HF model for the spin-isospin Landau-Midgal interaction $g^\prime$ which governs the response in the Gamow-Teller (GT) channel.
See \ref{appendix:LM} for more details. In the pure HF case, we simply have $g^\prime=0$, see Tab.~\ref{tab:lmg}.

\begin{table*}[t]
\tabcolsep=0.075cm
\def\arraystretch{2.0}
\caption{\label{tab:results_B}%
The parameters for the model B, fitted to the L1 mean value of a$_2$ and a$_4$, with the 3 values of $\kappa_{\rho}$. We also show the energy contribution of the various mesons at the Hartree and Fock levels. The delta meson contribution is negligible so it's not shown.}
\begin{tabular}{lcccc|cccc|cccccccc}
\hline
\multicolumn{5}{c|}{Parameters} & \multicolumn{4}{c}{NEP}  & \multicolumn{8}{|c}{Meson contribution to the binding energy} \\
\hline
model & $m_{s}$ & $g_{s}$ & $g_{\omega}$ & $C$ & $K_{\sat}$ & $E_{\sym}$ & $M_{D}^*/M_N$ & $E_\mathrm{K}$ & $E^{s}_\mathrm{H}$  & $E^{s}_\mathrm{F}$  & $E^{\omega}_\mathrm{H}$ & $E^{\omega}_\mathrm{F}$ & $E^{\pi}_\mathrm{F}$ & $E^{\rho^V}_\mathrm{F}$ & $E^{\rho^T}_\mathrm{F}$ & $E^{\rho^{VT}}_\mathrm{F}$ \\
 & MeV & & & & MeV & MeV & & MeV & MeV & MeV & MeV & MeV & MeV & MeV & MeV & MeV \\
\hline
RHF$_\mathrm{CC}(B,L1,\mathrm{NRT})$ & 932 & 14.175 & 8.688 & 1.72 & 324 & 27.93 & 0.72 & 18.9 & -97.1 & 12.7 & 73.3 & -13.8 & -5.6 & -4.6 & 0.0 & 0.0 \\
RHF$_\mathrm{CC}(B,L1,\mathrm{WRT})$ & 793 & 10.253 & 8.278 & 1.32 & 306 & 30.75 & 0.75 & 18.8 & -68.3 & 8.2 & 66.6 & -12.5 & -5.6 & -4.2 & -22.2 & 3.0 \\
RHF$_\mathrm{CC}(B,L1,\mathrm{SRT})$ & 332 & 1.80 & 9.244 & 0.26 & 280 & 46.78 & 0.75 & 17.5 & -11.1 & 0.9 & 83.0 & -15.3 & -5.6 & -5.1 & -87.5 & 7.1 \\
\hline
\end{tabular}
\end{table*}

There is however a practical issue induced by the pure $\pi$ and $\rho^T$ meson exchange, since these vertices contain derivative couplings which are known to generate a spurious repulsive contact term, as in Eq.~\eqref{eq:V_A}, see for instance Ref.~\cite{Bouyssy_1987} for more details. The effect of this spurious term is shown in Fig.~\ref{fig:V(q)} where we represent the interaction element $V_{A,\alpha}(\mathrm{q})$ ($\alpha=\rho^T$ and $\pi$), for the pure HF model by the blue solid line. The other cases will be discussed hereafter. The pure HF interaction shown in Fig.~\ref{fig:V(q)} for $\rho^T$ (top) and $\pi$ (bottom) mesons has an UV divergence: the pure HF interaction is not going to zero at large values of $\mathrm{q}$. In principle, this divergence does not lead to a collapse of the HF calculation since integrations over the momenta are limited to the Fermi momentum, but they have an impact over the short range properties of the interaction. In the following sections, we therefore address the question of short range properties of $\pi$ and $\rho^T$ vertex and we analyse different ways to treat them. We also study their impact on the values of the NEP.

\section{Improvement of the pure RHF calculation}
\label{sec:Improvment}

In this section we seek to improve on our previous calculation by treating the spurious repulsive contact term mentioned  before and by considering nucleons' finite size, and the SRC. These improvements affect the Fock terms treated in the previous section, and will be studied in three steps.

\subsection{Full removal of the spurious contribution from $\pi$ and $\rho^T$ mesons (model B)}
\label{sec:caseB}

A way to remove the spurious contact terms from  $\pi$ and $\rho^T$ mesons is to subtract the zero-rank part of the NN potential coming from the $\pi$ and Lorentz-tensor piece of the $\rho$ exchanges, hereafter called $\rho^T$~\cite{Bouyssy_1987}. This is equivalent to doing the following replacement in the Fock exchange terms:
\begin{equation}
V_{A,\alpha}(\mathrm{q}) \longmapsto V_{B,\alpha}(\mathrm{q})=\frac{\mathrm{q}^2}{\mathrm{q}^2 + m^2_\alpha} - 1 
= -\frac{m^2_\alpha}{\mathrm{q}^2 + m^2_\alpha}
\label{eq:caseB}
\end{equation}
We can see that the $\pi$ and $\rho^T$ exchange terms now take the form of an ordinary Yukawa potential and they give an attractive contribution to the energy per particle. This is the prescription suggested by the Orsay group, hereafter called the Orsay prescription. The other mesons are not significantly impacted by this prescription since this will only modify the values of their coupling constant~\cite{Bouyssy_1987}. Note that this prescription is most probably the simplest way to get rid of the spurious $\pi$ and $\rho^T$ terms when these meson fields are emitted by point-like nucleons. They are also employed in Refs.~\cite{Bernados_1993,Guichon_2006,Rikovska_Stone_2007}. In practice, the Orsay prescription precisely removes the contact terms $\Sigma_S^{E,c}$, $\Sigma_0^{E,c}$, and $\Sigma_V^{E,c}$ explicitly given in Eq.~\eqref{eq:sigma_S}, \eqref{eq:sigma_0} and \eqref{eq:sigma_V}. 

The impact of the Orsay prescription on the pure HF interaction is illustrated by the case $V_B$ shown in Fig.~\ref{fig:V(q)}: by removing the repulsive contact terms, the $\pi$ and $\rho^T$ Fock terms become purely attractive. The interaction elements are translated down as explicitly written in Eq.~\eqref{eq:caseB}. As a consequence, by curing the UV divergence, the Orsay prescription turns the $\pi$ and $\rho^T$ to be strongly attractive at long distance (small $\mathrm{q}$).

Similarly to the pure HF model where results in symmetric matter were given in Tab.~\ref{tab:results_A}, we investigate the matter properties for the model B in Tab.~\ref{tab:results_B}. The RH calculation being independent of the considered case, we refer to the results given in Tab.~\ref{tab:results_A} for RH$_\mathrm{CC}(L1)$. 

Let us first compare the models A and B for the NRT, where only the pion is considered. We observe that the isoscalar properties of nuclear matter are quite similar between these two cases: $m_s$, $g_s$, $g_\omega$, $C$, $K_\sat$, $M_D^*$. The symmetry energy $E_\sym$ is a bit increased in the model B compared to the model A and becomes closer to the empirical value.

We now analyse the impact of adding the $\rho^T$ interaction with different couplings: WRT and SRT. The scalar Dirac mass is not impacted by the different choices of $\rho^T$ coupling constant, but $K_\sat$ and $E_\sym$ are: $K_\sat$ goes down and get closer to the empirical expectation, while $E_\sym$ goes up and becomes even too large compared to the empirical value for the SRT coupling constant. The case with a strong $\rho^T$ (SRT) is however too extreme and shows the limitation of the model B, that we now discuss. As we have observed in Fig.~\ref{fig:V(q)}, the Orsay prescription changes the long-range properties of the $\pi$ and $\rho^T$ meson exchange potential, by turning them from repulsive and going to zero at long distance to be attractive and going to a finite value at long distance. As a consequence the contribution of these two mesons to the energy-density becomes increasingly large and negative as their coupling constant increases. The pion has an attractive contribution to the binding energy (it was repulsive in the model A), and the $\rho^T$ is also changed from repulsive in case A to attractive in case B. In fact, $\rho^T$ is very attractive for the two values of the $\rho^T$ coupling constant that we consider, contributing to about $-22$~MeV for WRT and to $-87$~MeV for SRT. For the model B, the SRT becomes extreme since the role of the scalar $s$ meson to the saturation of symmetric matter is almost quenched, being replaced by the $\rho^T$ term: saturation appears mostly as a balance between the attractive $\rho^T$ and the repulsive $\omega$ terms. The coupling constant $g_s$ in this case becomes abnormally small, and the symmetry energy is now predicted to be too large.

The spin-isospin Landau-Midgal interaction obtained for the model B can be deduced from Eq.~\eqref{eq:g_prime_B}.
Since $V_{B,\alpha}(\mathrm{q}\rightarrow 0)\rightarrow -1$, we obtain the values reported in Tab.~\ref{tab:lmg}.

In conclusion, the Orsay prescription for the derivative coupling terms induces a large contribution of the $\rho^T$ term, which produces an abnormal parameter set in the case of strong $\rho^T$ (SRT). It therefore leads to a preference for the low values of the $\rho^T$ coupling, which may be in contradiction with the expected SRT. The Orsay prescription is simple but have caveats, e.g. a large attraction of the derivative couplings at large distances. It is therefore preferable to investigate softer treatments of the short range properties of the nuclear interaction. Moreover, the models A and B ignore the nucleon finite-size, which induces naturally a cut-off for the high-$\mathrm{q}$ momenta. Nucleons are indeed composite particles, which are probed for the typical exchange energies in nuclear matter. This is the subject of the next section.

\subsection{Nucleon finite size effect (model C)}
\label{sec:caseC}

\begin{table*}[t]
\tabcolsep=0.075cm
\def\arraystretch{2.0}
\caption{\label{tab:results_C}%
The parameters for the model C, fitted to the L1 mean value of a$_2$ and a$_4$, with the 3 values of $\kappa_{\rho}$. We also show the energy contribution of the various mesons at the Hartree and Fock levels. The delta meson contribution is negligible so it's not shown.}
\begin{tabular}{lcccc|cccc|cccccccc}
\hline
\multicolumn{5}{c|}{Parameters} & \multicolumn{4}{c}{NEP}  & \multicolumn{8}{|c}{Meson contribution to the binding energy} \\
\hline
model & $m_{s}$ & $g_{s}$ & $g_{\omega}$ & $C$ & $K_{\sat}$ & $E_{\sym}$ & $M_{D}^*/M_N$ & $E_\mathrm{K}$ & $E^{s}_\mathrm{H}$  & $E^{s}_\mathrm{F}$  & $E^{\omega}_\mathrm{H}$ & $E^{\omega}_\mathrm{F}$ & $E^{\pi}_\mathrm{F}$ & $E^{\rho^V}_\mathrm{F}$ & $E^{\rho^T}_\mathrm{F}$ & $E^{\rho^{VT}}_\mathrm{F}$ \\
 & MeV & & & & MeV & MeV & & MeV & MeV & MeV & MeV & MeV & MeV & MeV & MeV & MeV \\
\hline
RHF$_\mathrm{CC}(C,L1,\mathrm{NRT})$  & 981 & 15.711 & 8.098 & 1.86 & 331 & 22.15 & 0.74 & 19.3 & -109.1 & 12.7 & 63.7 & -10.5 & 11.3 & -3.5 & 0.0 & 0.0 \\
RHF$_\mathrm{CC}(C,L1,\mathrm{WRT})$ & 948 & 14.646 & 7.163 & 1.76 & 319 & 19.93 & 0.77 & 19.5 & -101.1 & 11.5 & 49.8 & -8.3 & 11.3 & -2.8 & 1.9 & 1.9 \\
RHF$_\mathrm{CC}(C,L1,\mathrm{SRT})$ & 920 & 13.792 & 6.275 & 1.68 & 308 & 18.20 & 0.80 & 19.9 & -95.0 & 10.6 & 38.2 & -6.4 & 11.4 & -2.2 & 4.7 & 2.6 \\
\hline
\end{tabular}
\end{table*}

We now explore the modifications for the nuclear interaction induced by the inclusion of nucleon finite-size. This is done by introducing form factors to each meson-nucleon vertex, and we consider, for simplicity, the monopole form factors  (FFs) prescription, which reads for the meson $\alpha$,
\begin{equation}
F_\alpha(\mathrm{q}) = \left(\frac{\Lambda_\alpha^2}{\mathrm{q}^2 + \Lambda_\alpha^2}\right) \, ,
\label{eq:ff}
\end{equation}
where $\Lambda_\alpha$ may vary with meson $\alpha$.
In the following, we associate a FF to all the mesons fields, e.g., $s$, $\delta$, $\omega$, $\rho$, $\pi$.
The propagator is modified as follows:
\begin{equation}
\label{eq:propagatorC}
D_\alpha(\mathrm{q}) \longmapsto D_\alpha(\mathrm{q})=\frac{1}{\mathrm{q}^2 + m_\alpha^2}\left(\frac{\Lambda_\alpha^2}{\mathrm{q}^2 + \Lambda_\alpha^2}\right)^2 \, ,
\end{equation}
for mesons $\alpha=s$, $\delta$, $\omega$, $\rho$, and $\pi$.
Note however that for the interaction $\rho^T$ and for the meson $\pi$ there are additional contact terms, once we rewrite the interaction element as in Eq.~\eqref{eq:V_A}, which will be detailed hereafter. Note also that for the interaction $\rho^{VT}$, the propagator becomes
\begin{equation}
\label{eq:V_C_mixte}
 D_{\rho^{VT}}(\mathrm{q})=\frac{1}{\mathrm{q}^2 + m_\rho^2}\left(\frac{\Lambda_{\rho^V}^2}{\mathrm{q}^2 + \Lambda_{\rho^V}^2}\right) \left(\frac{\Lambda_{\rho^T}^2}{\mathrm{q}^2 + \Lambda_{\rho^T}^2}\right) \, ,
\end{equation}
since it has both vertices from $\rho^V$ and $\rho^T$ which are different.
Note that our approach is similar to the one suggested in Ref.~\cite{Toki2010}, where the following FF is considered,
\begin{equation}
F_\alpha^\prime(\mathrm{q}) = \left(\frac{\Lambda_\alpha^2-m_\alpha^2}{\mathrm{q}^2+\Lambda_\alpha^2} \right) \, .
\label{eq:ff2}
\end{equation}
This FF is the same as the one considered in the Bonn potential~\cite{Machleidt:1987}.
The two FF only differ by a global constant,
$F_\alpha=(1-m_\alpha^2/\Lambda_\alpha^2)F_\alpha^\prime$.
We employ $F_\alpha$ since $F_\alpha\rightarrow 1$ as $\mathrm{q}\rightarrow 0$: at low momentum transfer ($\mathrm{q}\ll\Lambda_\alpha,m_\alpha$), the interaction is not impacted by nucleon finite size.

By doing a simple element decomposition in Eq.~\eqref{eq:propagatorC}, we obtain
\begin{subequations}
\begin{eqnarray}
D_\alpha(\mathrm{q})&=& \left(\frac{\Lambda_\alpha^2}{\Lambda_\alpha^2 - m_\alpha^2}\right)^2 
\Big[\frac{1}{\mathrm{q}^2 + m_\alpha^2} - \frac{1}{\mathrm{q}^2 + \Lambda_\alpha^2} \nonumber \\ 
&&\hspace{1cm}+(\Lambda_\alpha^2-m_\alpha^2)\frac{d}{d\Lambda_\alpha^2}\frac{1}{\mathrm{q}^2 + \Lambda_\alpha^2} \Big], \\
D_{\rho^{VT}}(\mathrm{q}) &=& \frac{\Lambda_{\rho^V}^2 \Lambda_{\rho^T}^2}{(\Lambda_{\rho^V}^2-m_\rho^2)(\Lambda_{\rho^T}^2-m_\rho^2)} \Big [\frac{1}{\mathrm{q}^2 + m_\rho^2} \nonumber \\ &&-\frac{\Lambda_{\rho^T}^2-m_\rho^2}{\Lambda_{\rho^T}^2-\Lambda_{\rho^V}^2}\frac{1}{\mathrm{q}^2 + \Lambda_{\rho^V}^2} \nonumber \\
&&+ \frac{\Lambda_{\rho^V}^2-m_\rho^2}{\Lambda_{\rho^T}^2-\Lambda_{\rho^V}^2}\frac{1}{\mathrm{q}^2 + \Lambda_{\rho^T}^2} \Big] \, .
\end{eqnarray}
\end{subequations}
We replace the functions $\theta_\alpha$ and $\phi_\alpha$, see Eqs.~\eqref{eq:theta} and \eqref{eq:phi}, by
$\theta_\alpha^{FF}$ and $\phi_\alpha^{FF}$ defined as,
\begin{subequations}
\begin{eqnarray}
\theta^{FF}_\alpha &=& \left(\frac{\Lambda_\alpha^2}{\Lambda_\alpha^2 - m^2_\alpha}\right)^2\Big[\theta_\alpha - \theta_\alpha(m_\alpha=\Lambda_\alpha) \nonumber \\
&&\hspace{0.5cm}+(\Lambda_\alpha^2-m_\alpha^2)\frac{d}{d\Lambda_\alpha^2}\theta_\alpha(m_\alpha=\Lambda_\alpha)\Big], \label{eq:theta_FF} \\ 
\phi^{FF}_\alpha &=& \left(\frac{\Lambda_\alpha^2}{\Lambda_\alpha^2 - m_\alpha^2}\right)^2\Big[\phi_\alpha - \phi_M(m_\alpha=\Lambda_\alpha) \nonumber \\
&&\hspace{0.5cm}+ (\Lambda_\alpha^2-m_\alpha^2)\frac{d}{d\Lambda_\alpha^2}\phi_M(m_\alpha=\Lambda_\alpha)\Big], \label{eq:phi_FF}
\end{eqnarray}
\end{subequations}

\begin{subequations}
\begin{eqnarray}
\theta^{FF}_{\rho^{VT}} &=& \frac{\Lambda_{\rho^V}^2 \Lambda_{\rho^T}^2}{(\Lambda_{\rho^V}^2-m_\rho^2)(\Lambda_{\rho^T}^2-m_\rho^2)} \Big [\theta_\rho \nonumber \\ && -\frac{\Lambda_{\rho^T}^2-m_\rho^2}{\Lambda_{\rho^T}^2-\Lambda_{\rho^V}^2} \theta_\rho(m_\rho=\Lambda_{\rho^V}) \nonumber \\
&& + \frac{\Lambda_{\rho^V}^2-m_\rho^2}{\Lambda_{\rho^T}^2-\Lambda_{\rho^V}^2}\theta_\rho(m_\rho=\Lambda_{\rho^T})], \\
\phi^{FF}_{\rho^{VT}} &=& \frac{\Lambda_{\rho^V}^2 \Lambda_{\rho^T}^2}{(\Lambda_{\rho^V}^2-m_\rho^2)(\Lambda_{\rho^T}^2-m_\rho^2)} \Big [\phi_\rho \nonumber \\ && -\frac{\Lambda_{\rho^T}^2-m_\rho^2}{\Lambda_{\rho^T}^2-\Lambda_{\rho^V}^2} \phi_\rho(m_\rho=\Lambda_{\rho^V}) \nonumber \\
&& + \frac{\Lambda_{\rho^V}^2-m_\rho^2}{\Lambda_{\rho^T}^2-\Lambda_{\rho^V}^2}\phi_\rho(m_\rho=\Lambda_{\rho^T})].
\end{eqnarray}
\end{subequations}
The contact self-energies \eqref{eq:selfSc}-\eqref{eq:selfVc} for $\rho^T$ and $\pi$ channels are obtained from the following transformation
\begin{equation}
4\mathrm{p}\mathrm{p}^\prime \longmapsto -\Lambda_\alpha^4\frac{d}{d\Lambda_\alpha^2}\theta_\alpha(\Lambda_\alpha)
\end{equation}
for $\Sigma_S^{E,\mathrm{c}}$ and $\Sigma_0^{E,\mathrm{c}}$, while for $\Sigma_V^{E,\mathrm{c}}$ we have
\begin{eqnarray}
\Sigma_V^{E,\pi,\mathrm{c},FF} &=& \left(\frac{g_A}{2f_\pi}\right)^2 \!\!
\langle \Lambda_\alpha^4\frac{d}{d\Lambda_\alpha^2}\phi_\alpha(m_\alpha=\Lambda_\alpha) 2\hat{P}(\mathbf{p}^\prime) \rangle \, , \nonumber\\
\Sigma_V^{E,\rho^T,\mathrm{c},FF} &=& \left(\frac{f_{\rho}}{2M}\right)^2 \!\!
\langle \Lambda_\alpha^4\frac{d}{d\Lambda_\alpha^2}\phi_\alpha(m_\alpha=\Lambda_\alpha) 2\hat{P}(\mathbf{p}^\prime) \rangle \nonumber \, .
\end{eqnarray}

In the following, the self-energies including nucleon finite size through the FF are named as 
$\Sigma^{FF}$.

Note that at low momentum ($\mathrm{p},\mathrm{p}^\prime\ll\Lambda_\alpha,m_\alpha$), we recover the results of the pure HF case (model A), see 
sec.~\ref{sec:caseA}, since
\begin{eqnarray}
&&\theta^{FF}_\alpha\rightarrow \theta_\alpha, \,\, \quad
\Lambda_\alpha^4\frac{d}{d\Lambda_\alpha^2}\theta_\alpha(m_\alpha=\Lambda_\alpha)\rightarrow-4\mathrm{p}\mathrm{p}^\prime, \\
&&\phi^{FF}_\alpha\rightarrow \phi_\alpha, \,\, \quad
\Lambda_\alpha^4\frac{d}{d\Lambda_\alpha^2}\phi_\alpha(m_\alpha=\Lambda_\alpha)\rightarrow 0 \, .
\end{eqnarray}

It is interesting to remark that the FF depends on the cut-off $\Lambda_\alpha$, which can be different for different channels since nucleon finite size is seen differently for various probes. For the practical results shown in this paper and for simplicity, we consider $\Lambda_\alpha$ = 1 GeV for all mesons $\alpha$, except for $\rho^T$ for which $\Lambda_{\rho^T}$ = 2 GeV, which is needed to obtain a sufficiently large value of the spin-isospin Landau-Migdal $g'$ parameter. Our FF are comprised between 1 and 2 GeV, see Tab.~\ref{tab:fixedparams}, as in the Bonn potential~\cite{Machleidt:1987}.

The interaction element $V_C(\mathrm{q})$, defined as
\begin{equation}
\label{eq:V_C}
V_{C,\alpha}(\mathrm{q}) = V_{A,\alpha}(\mathrm{q}) \left( F_\alpha(\mathrm{q})\right)^2, \quad (\alpha = \pi, \rho^T)
\end{equation}
is shown in Fig.~\ref{fig:V(q)}. At low momentum transfer ($\mathrm{q}\ll\Lambda_\alpha,m_\alpha$) we have 
$V_C(\mathrm{q})=V_A(\mathrm{q})$ as expected, while at high momentum transfer, we now obtain the limit $V_C\rightarrow 0$. The inclusion of the nucleon finite size therefore regularizes the UV divergence discussed in Sec.~\ref{sec:caseA}. 
The FF treatment is therefore equivalent, from a mathematical view point, to the Orsay prescription, but without changing the low momentum limit of the interaction element $V(\mathrm{q})$. The physical origin of the FF is however very different from the Orsay prescription: The first is due to the nucleon finite size, while the latter is supposed to be generated by the SRC. We will consider both physical effects, nucleon finite size and SRC, in the following section.

We show in Tab.~\ref{tab:results_C} our results for case C in symmetric matter and at saturation  density. Since the Fermi momentum $p_F$ at saturation density is about $p_F\sim 270$~MeV, the low momentum approximation applies for all mesons, except for the $\pi$ channel. As a consequence, results shown in Tab.~\ref{tab:results_C} are very similar to the pure HF model, see Tab.~\ref{tab:results_A}, except for the $\pi$ channel. The change is however small in the $\pi$ channel: We obtain a contribution of the $\pi$ of about 13.5~MeV in Tab.~\ref{tab:results_A}, which is about 13.4~MeV in Tab.~\ref{tab:results_C}. It induces a slight rearrangement of the coupling constants, but the overall effect remains moderate. The impact of the FF is expected to appear more clearly as the density increases. However, let us note that Fermi momenta of about 500~MeV are expected for densities of about $8n_\sat$. The low density limit therefore applies well, even for the largest density we may consider, except again, for the pion.

The finite-size effect does not change the pure HF spin-isospin Landau-Midgal parameter since $F_\alpha(\mathrm{q}\rightarrow 0)\rightarrow 1$. We then obtain $g^\prime_C=g^\prime_A=0$, see Tab.~\ref{tab:lmg}.

In conclusion, the FF treatment resolves the UV divergence of the interaction element originating from the derivative couplings, while our results including FF remain very close to the pure HF model shown in Sec.~\ref{sec:caseA}. In particular, we obtain an incompressibility modulus $K_\sat$ a bit higher than the phenomenological one, and a symmetry energy $E_\sym$ a bit lower. The interaction element originating from derivative couplings are known to generate a strong interaction at short distance, even after the FF regularisation~\cite{Toki2010}. As a consequence, SRC are expected to be non negligible. In the next section, we therefore include SRC by the use of Jastrow ansatz.

\subsection{Nucleon finite size and short range correlations (model D)}
\label{sec:caseD}


\begin{table*}[t]
\tabcolsep=0.075cm
\def\arraystretch{2.0}
\caption{\label{tab:results_D}%
The parameters for the model D, fitted to the L1 mean value of a$_2$ and a$_4$, with the 3 values of $\kappa_{\rho}$. We also show the energy contribution of the various mesons at the Hartree and Fock levels. The delta meson contribution is negligible so it's not shown.}
\begin{tabular}{lcccc|cccc|cccccccc}
\hline
\multicolumn{5}{c|}{Parameters} & \multicolumn{4}{c}{NEP}  & \multicolumn{8}{|c}{Meson contribution to the binding energy} \\
\hline
model & $m_{s}$ & $g_{s}$ & $g_{\omega}$ & $C$ & $K_{\sat}$ & $E_{\sym}$ & $M_{D}^*/M_N$ & $E_\mathrm{K}$ & $E^{s}_\mathrm{H}$  & $E^{s}_\mathrm{F}$  & $E^{\omega}_\mathrm{H}$ & $E^{\omega}_\mathrm{F}$ & $E^{\pi}_\mathrm{F}$ & $E^{\rho^V}_\mathrm{F}$ & $E^{\rho^T}_\mathrm{F}$ & $E^{\rho^{VT}}_\mathrm{F}$ \\
 & MeV & & & & MeV & MeV & & MeV & MeV & MeV & MeV & MeV & MeV & MeV & MeV & MeV \\
\hline
RHF$_\mathrm{CC}(D,L1,\mathrm{NRT})$ & 965 & 15.200 & 8.176 & 1.81 & 328 & 23.2 & 0.74 & 19.3 & -105.2 & 12.1 & 64.9 & -10.8 & 7.0 & -3.6 & 0.0 & 0.0 \\
RHF$_\mathrm{CC}(D,L1,\mathrm{WRT})$ & 903 & 13.296 & 7.457 & 1.64 & 314 & 22.9 & 0.77 & 19.4 & -90.8 & 10.1 & 54.0 & -9.0 & 7.0 & -3.0 & -5.9 & 2.1 \\
RHF$_\mathrm{CC}(D,L1,\mathrm{SRT})$ & 819 & 10.943 & 6.932 & 1.40 & 296 & 24.7 & 0.80 & 19.6 & -73.7 & 7.8 & 46.7 & -7.8 & 7.1 & -2.6 & -16.4 & 3.2 \\
\hline
\end{tabular}
\end{table*}

In this section, we consider both the nucleon finite size, with FF, and the short-range correlations (SRC) in a way which is more general than the Orsay prescription, as it is detailed in the model B discussed in Sec.~\ref{sec:caseB}. Note that we introduce SRC only for the $\rho^T$ and $\pi$ interactions for practical reasons, but in principle, all mesons exchange potentials could potentially carry corrections at short distance from SRC, but not as impactful. In addition, it is natural to introduce the SRC in coordinate space since they impact the interaction element at short distance. In the following, we introduce SRC as a modification of the interaction element by replacing $D({\bf r}-{\bf r'})$ in coordinate space by $D({\bf r}-{\bf r'})\left[1-G({\bf r}-{\bf r'}) \right]$,
where
$G({\bf r}-{\bf r'})$ is a two body correlation function which forbids the presence of two nucleons at the same point, by imposing the following property $G({\bf r}={\bf r'})=1$. 
The two body correlation function $G(r)$ is identified with the Jastrow function, $G(r) \equiv j_0(\mathrm{\mathrm{q_c}}r)$, with $j_0$ the Bessel function of the first kind and $\mathrm{q_c}$ a parameter controlling the coordinate shape of the correlation function. This is a similar approach to ones in Refs.~\cite{MARCOS1996,Hu2011} in which a more general two body correlation function is considered, which could potentially depend on more than one parameter, rather than the one parameter Bessel function we used.

The Fourier transformation of the derivative coupling terms reads
\begin{equation}
\partial_i\partial'_jD({\bf r}-{\bf r'}) = \int\!\!\frac{d {\bf q}}{(2\pi)^3} q^iq^jD(\mathrm{q})e^{i{\bf q}\cdot({\bf r}-{\bf r'})}
\end{equation}
and the Fourier transform of this correlation function (defined from the Jastrow ansatz) is
\begin{equation}
G(q) = \int\!\!\frac{d {\bf r}}{(2\pi)^3}e^{-i{\bf q}\cdot {\bf r}}G(r) = \frac{1}{4\pi \mathrm{q_c}^2}\delta(\mathrm{q}-\mathrm{\mathrm{q_c}})
\end{equation}

From Eq.~\eqref{eq:exchange_hamiltonian}, 
we have for the $\rho^T$ and $\pi$ mesons couplings, after convolution with the Jastrow function,
\begin{eqnarray}
\partial_i\partial'_j D_\alpha({\bf r}-{\bf r'}) &=& \int\!\!\frac{d {\bf k}}{(2\pi)^3}e^{i{\bf k}\cdot({\bf r}-{\bf r'})}k_ik_j D_\alpha(\mathrm{k}) \nonumber \\
&& \hspace{0.5cm}\times \int\!\!d {\bf t}e^{i{\bf t}\cdot({\bf r}-{\bf r'})}\left[\delta^{(3)}(t)-G(t) \right] \nonumber \\
&=& \int\!\!\frac{d {\bf q}}{(2\pi)^3}e^{i{\bf q}\cdot({\bf r}-{\bf r'})}\nonumber \\
&& \hspace{-2cm} \times
\Big[q_iq_j D_\alpha(\mathrm{q})
-\int\!\!d {\bf k}D_\alpha(\mathrm{k})G({\bf q}-{\bf k})k_ik_j \Big] \, .
\end{eqnarray}
We can write the last term on the right, the correlation term, from symmetry principles, as a function of $q_iq_j$ and of $\delta_{ij}$:
\begin{equation}
\int\!\!d {\bf k}D_\alpha(\mathrm{k})G({\bf q}-{\bf k})k_ik_j = g'_\alpha(\mathrm{q})\delta_{ij} + (3\hat{q}_i\hat{q}_j - \delta_{ij})h'_\alpha(\mathrm{q})
\end{equation}
with
\begin{eqnarray}
g'_\alpha(\mathrm{q}) &=& \frac{1}{3}\int\!\!d {\bf k}\mathrm{k}^2D_\alpha(\mathrm{k})G({\bf q}-{\bf k}) \\
(g'_\alpha+2h'_\alpha)(\mathrm{q}) &=&  \int\!\!d {\bf k}({\bf k}\cdot\tilde{q})^2D_\alpha(\mathrm{k})G({\bf q}-{\bf k})
\end{eqnarray}
Using the Oset-Toki-Weise (OTW) approximation (and fixing $\omega=0$ since we consider the static approximation) from Ref.~\cite{OSET1982281}, we obtain
\begin{subequations}
\begin{align}
g'_\alpha(\mathrm{q}) &= \frac{1}{3}\frac{\mathrm{q}^2 + \mathrm{q_c}^2}{\mathrm{q}^2+m_\alpha^2+\mathrm{q_c}^2}, \\
h'(\mathrm{q}) &= \frac{1}{3}\frac{\mathrm{q}^2}{\mathrm{q}^2+m_\alpha^2+\mathrm{q_c}^2}   
\end{align}
\end{subequations}
Finally, the interaction element is expressed as
\begin{eqnarray}
\partial_i\partial'_j D_\alpha({\bf r}-{\bf r'}) &=& \int\!\!\frac{d {\bf q}}{(2\pi)^3}e^{i{\bf q}\cdot({\bf r}-{\bf r'})} \nonumber \\
&&\hspace{-3cm}\times \Big[ \hat{q}_i\hat{q}_j\left(D_\alpha(\mathrm{q})-3h'_\alpha(\mathrm{q})\right) 
-\delta_{ij}\left(g'_\alpha(\mathrm{q})-h'_\alpha(\mathrm{q})\right) \Big] \, .
\end{eqnarray}

We can now proceed to recalculate the $\pi$ and $\rho^T$ interaction, with the corresponding expressions appearing in \ref{app:selfD}.
The interaction element including the FF is modified as described in Eq.~\eqref{eq:V_C},
as for the terms with $g'$ and $h'$, the form factor will become $F^2(\mathrm{q}) \longmapsto F^2(\mathrm{q}^2 \longmapsto \mathrm{q}^2+ \mathrm{q_c}^2)$.

We will now detail the case of the $\pi$ meson, from which one could also deduce the case of $\rho^T$. One has~:
\begin{align}
\label{eq:Sig0_FF_SRC}
&\Sigma^\mathrm{\pi,SRC-FF}_0 = \frac{2}{(4\pi)^2}\left(\frac{g_A}{2f_\pi}\right)^2\sum\limits_{\tau'}\tau_{\pi}^2 \int_{0}^{k_F}\mathrm{p'}^2d\mathrm{p'}~ d(\cos\theta) 
 \nonumber \\ 
& \times \left(\mathrm{q}^2D_\pi(\mathrm{q})F^2_{\pi}(\mathrm{q}) - \frac{\mathrm{q}^2+\mathrm{q_c}^2}{\mathrm{q}^2+\mathrm{q_c}^2+m^2_{\pi}}F^2_{\pi}(\sqrt{\mathrm{q}^2+\mathrm{\mathrm{\mathrm{q_c}}}^2}) \right)
\end{align}
which reduces to
\begin{equation}
\label{eq:sigma_0_SRC}
\Sigma^{SRC-FF}_0(\mathrm{p},\tau) = \Sigma^{FF}_0(\mathrm{q}) - \Sigma^{FF}_0(\mathrm{q}^2 \longmapsto \mathrm{q}^2+\mathrm{q_c}^2).
\end{equation}

We can then define our new interaction element for the case D as
\begin{equation}
\label{eq:V_D}
V_D(\mathrm{q})=V_C(\mathrm{q})-V_C(\mathrm{q}\rightarrow\sqrt{\mathrm{q}^2+\mathrm{q}_c^2})
\end{equation}

The second piece of Eq.~\eqref{eq:sigma_0_SRC} can be easily deduced by replacing the functions $\theta_\alpha$ and $\phi_\alpha$ in  Eq.~\eqref{eq:theta_FF} and \eqref{eq:phi_FF} by:
\begin{subequations}
\begin{eqnarray}
\theta_\alpha(m_\alpha) &\longmapsto& \theta_\alpha(m_\alpha \longmapsto \sqrt{m_\alpha^2 + \mathrm{q_c}^2}), \\
\theta_\alpha(\Lambda_\alpha) &\longmapsto& \theta_\alpha(\Lambda_\alpha \longmapsto \sqrt{\Lambda_\alpha^2 + \mathrm{q_c}^2}), \\
\phi_\alpha(m_\alpha) &\longmapsto& \phi_\alpha(m_\alpha \longmapsto \sqrt{m_\alpha^2 + \mathrm{q_c}^2}), \\
\phi_\alpha(\Lambda_\alpha) &\longmapsto& \phi_\alpha(\Lambda_\alpha \longmapsto \sqrt{\Lambda_\alpha^2 + \mathrm{q_c}^2}).
\end{eqnarray}
\end{subequations}
The same applies to $\Sigma_S$ which has the same integrals appearing. We now use $\theta_\alpha^{FF,SRC}$ and $\phi_\alpha^{FF,SRC}$ for the functions modified as mentioned above. We can turn now to $\Sigma_V$ which has an extra step to consider:
\begin{align}
&\Sigma^{\pi,SRC-FF}_V = -\frac{2}{(4\pi)^2}\left(\frac{g_A}{2f_\pi}\right)^2\sum\limits_{\tau'}\tau_{\pi}^2 \nonumber \\
& \times \int_{0}^{k_F}\mathrm{p'}^2 d\mathrm{p'} ~d(\cos\theta)  \times \nonumber \\
& \Big[ \big(\mathrm{q}^2D_\pi(\mathrm{q})F^2_{\pi}(\mathrm{q}) - \mathrm{q}^2D_{\pi}(\sqrt{\mathrm{q}^2+\mathrm{q}_c^2})F^2_{\pi}(\sqrt{\mathrm{q}^2+\mathrm{q}_c^2}) \big)\times \nonumber \\
& \big(\hat{P}(\mathrm{p}^\prime)\cos\theta-\frac{2}{\mathrm{q}^2}(\tilde{\bf p}\cdot{\bf q})(\hat{P}(\bf p^\prime)\cdot{\bf q})\big) \nonumber \\
& - \frac{1}{3}\mathrm{q_c}^2D_{\pi}(\sqrt{\mathrm{q}^2+\mathrm{q}_c^2})F^2_{\pi}(\sqrt{\mathrm{q}^2+\mathrm{q}_c^2})  \hat{P}(\mathrm{p}^\prime)\cos\theta \Big]
\end{align}

We have three parts in the previous expression: The first term with
$\mathrm{q}^2V_{\pi}(\mathrm{q})F^2_{\pi}(\mathrm{q})$ gives $\Sigma_V^{FF}$. \\
The second term with $\mathrm{q}^2V_{\pi}(\sqrt{\mathrm{q}^2+\mathrm{q}_c^2})F^2_{\pi}(\sqrt{\mathrm{q}^2+\mathrm{q}_c^2})$ is also identified to $\Sigma_V^{FF}$, but for the case of the $\rho^T$ one, we should also replace the function $C_{\rho^T}$ of Tab.~\ref{tab:coefficients} as follows: 
\begin{eqnarray}
&&C_{\rho^T} = 4\left(\frac{\emph{f}_{\rho}}{2M}\right)^2m^2_{\rho}\nonumber\\
&&\times \left [(\mathrm{p}^2+\mathrm{p}^{\prime 2}-(m^2_{\rho} + \mathrm{q^2_c})/2)\phi^{FF,SRC}_{\rho} - \mathrm{p}\mathrm{p}^\prime\theta^{FF,SRC}_{\rho}\right ] \nonumber\\
\end{eqnarray}
We are left with the last term which is re-expressed as,
\begin{eqnarray} 
\left(\frac{g_A}{2f_\pi}\right)^2 \langle \frac{2}{3}\mathrm{q^2_c} \hat{P}(\mathrm{p}^\prime) \phi^{FF,SRC} \rangle .
\end{eqnarray}

\begin{figure}[t]
\centering
\includegraphics[width=0.5\textwidth]{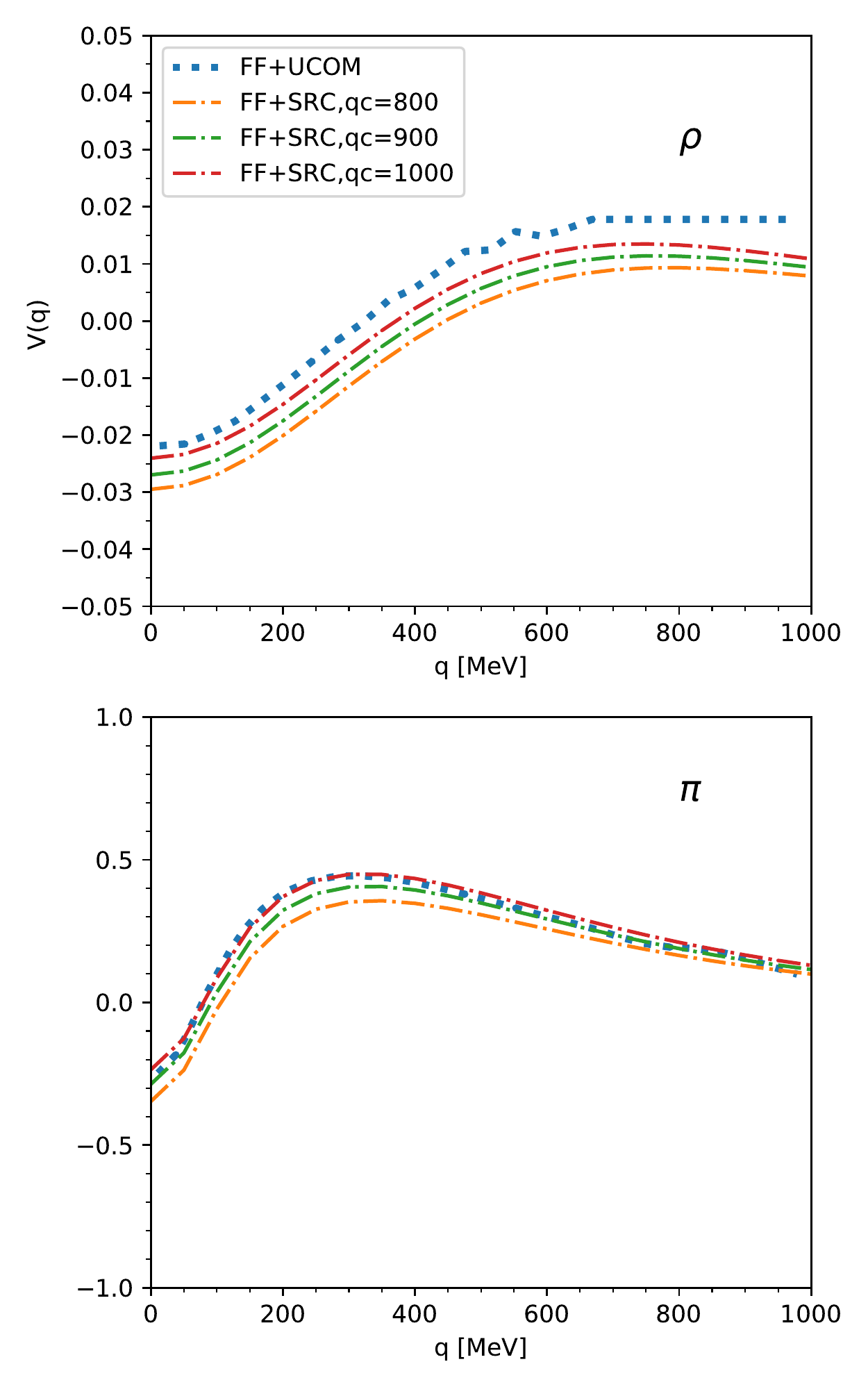}
\caption{The tensor $\rho$ and $\pi$ meson interaction in the case of the UCOM method (see Ref.~\cite{Toki2010}) compared to $\tilde{V}_D$ in the case of the Jastrow function. We vary the value of q$_c$ until we reproduce the same interaction, corresponding to a value of q$_c$ = 1~GeV. The FF were all taken to have the same form and the same cut-off as in Ref.~\cite{Toki2010}}.
\label{fig:qc}
\end{figure}

In this way, the short range properties of the interaction in the $\pi$ and $\rho$-tensor channels can be optimized by fixing the parameter $\mathrm{q}_c$. For large values of $\mathrm{q}_c\gg m_\alpha$, the Jastrow treatment of the SRC reproduces the Orsay prescription (in the abscence of FF), discussed in section \ref{sec:caseB}. The Jastrow approach allows however to incorporate SRC in a smoother way by varying the parameter $\mathrm{q}_c$. In the present study, we fix the value of the parameter $\mathrm{q}_c$ to reproduce the microscopic prediction for the interaction element $V(\mathrm{q})$ based on the UCOM approach~\cite{Toki2010}. This is illustrated in Fig.~\ref{fig:qc} where we compare the UCOM prediction for the $\rho^T$ and $\pi$ meson exchange to a Jastrow approximation of the same interaction element for values of $\mathrm{q}_c=800$, 900 and 1000 MeV. Note that in this case, we have employed the same FF as in Ref.~\cite{Toki2010}, and so we use $\tilde{V}_D(\mathrm{q})=V_D(\mathrm{q})/(1-m_\alpha^2/\Lambda_\alpha^2)^2$, see Eq.~\eqref{eq:ff2}. In the following, we fix $\mathrm{q}_c=1000$~MeV, see Tab.~\ref{tab:fixedparams}, since it reproduces very well the microscopic prediction from the UCOM approach for the two channels $\rho^T$ and $\pi$.

Looking at Tab.~\ref{tab:results_D}, we can see that $m_{s}$ and $g_{s}$ have decreased, and for $\kappa_{\rho} = 6.6 $ we are within the L$\sigma$M value of these parameters. We can see a more "moderate" value for the $\rho^T$ interaction with a value of -16 MeV, and the pion contribution has also been cut-down to 7 MeV. 

The dimensionless spin-isospin Landau-Midgal parameter for the model D is given in Eq.~\eqref{eq:g_prime_D} and Tab.~\ref{tab:lmg}. As mentioned in the \ref{appendix:LM}, considering the SRC for all interaction channels in addition to the derivative couplings considered so far, we would have obtained $g^\prime_D = 0.08/0.31/0.61$ for the following choice of the $\rho^T$ coupling: NRT/WRT/SRT. These values are close to $g^\prime\sim 0.6-0.7$, which was obtained in the analysis of the transfer reaction $(n,p)$ and $(p,n)$ at 300-500~MeV~\cite{Ichimura2006}. The Landau parameter $g^\prime$ can also be deduced from low-energy nuclear resonances, such as the spin and spin–isospin collective modes, e.g. magnetic dipole (M1) and Gamow–Teller (GT) states. The "nuclear" Landau parameter $\tilde{G}^\prime$, note the different normalisation of this parameter in hadron and in nuclear physics inducing $g^\prime \approx \tilde{G}^\prime/1.6$, see \ref{appendix:LM}, has been deduced from the analysis of the GT mode: a model based on Woods–Saxon single-particle states plus one-pion and rho meson exchange interactions gives $\tilde{G}^\prime = 1.3 \pm 0.2$, see Refs.~\cite{Osterfeld1992,Suzuki1999,Margueron2009} and references therein. A slightly different value, $\tilde{G}^\prime = 1.0 \pm 0.1$, was derived from observed GT and M1 strength distributions using the phenomenological energy density functionals DF3 \cite{Borzov1984,Borzov2006}. These values are consistent with the values obtained for $g^\prime_D$.

\begin{table}[t]
\tabcolsep=0.9cm
\def\arraystretch{1.3}
\caption{Parameters fixing the SRC and nuclear finite size in models C and D.}
\label{tab:fixedparams}
\begin{tabular}{ccc}
\hline
$\mathrm{q}_c$  & $\Lambda_\alpha$ & $\Lambda_{\rho^T}$ \\
\hline
1 GeV & 1 GeV & 2 GeV \\
\hline
\end{tabular}
\end{table}

We finally discuss the NEP. The RH$_\mathrm{CC}$ model predict the incompressibility modulus to be around 265~MeV, with an uncertainty estimated to be about 15~MeV~\cite{Rahul2022}. It is compatible with the experimental value 230$\pm $20~MeV~\cite{Margueron2018}.
In the RHF$_\mathrm{CC}$ models however, by adding the contribution of the Fock terms, the incompressibility modulus is increased to about 306-326~MeV in the pure HF model (model A), see Sec.~\ref{sec:caseA}, and 296-328~MeV once the SRC and the FF are taken into account. The uncertainties in the incompressibility modulus reported here is the one coming only from the uncertainty in the $\rho^T$ coupling constant. The incompressibility is therefore still above the experimental one by about 20\%.
Concerning the symmetry energy, RH$_\mathrm{CC}$ model predict 18.05~MeV, with an uncertainty of about 1~MeV~\cite{Rahul2022}, the pure HF model predict an increase of the symmetry energy, spanning from 19.7 to 23.7~MeV and our ultimate model with SRC and FF predict it to be about 23.2-24.7~MeV. The inclusion of the Fock term therefore lead to an increase of the symmetry energy by about 30\%. We then conclude that there is an  improvement, although moderate, of the prediction for two NEP which are not used in the fitting protocol ($E_\sym$ and $K_\sat$) in our ultimate model compared to the one at the Hartree approximation.

\begin{figure*}[t]
\centering
\includegraphics[width=0.8\textwidth]{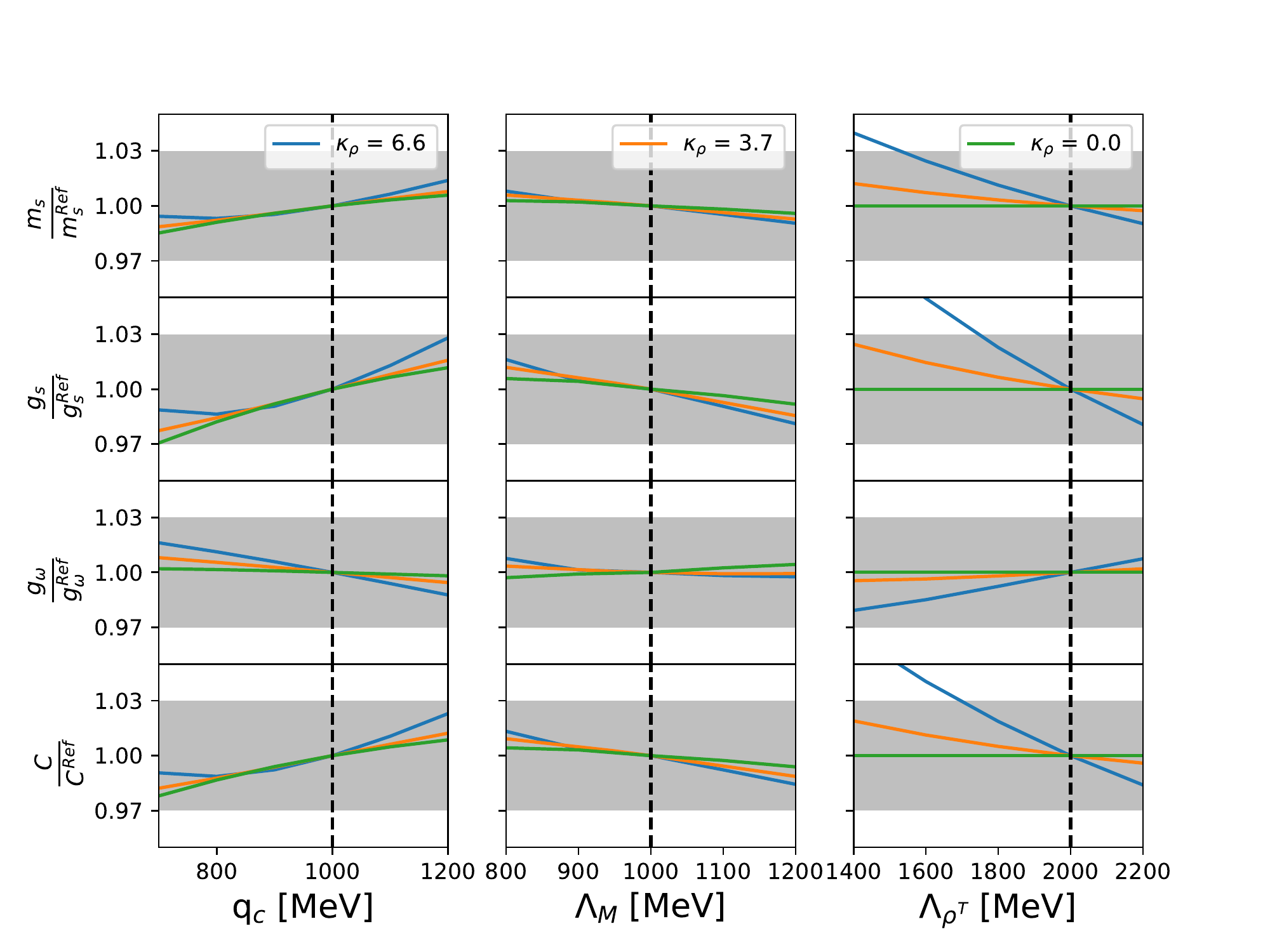}
\caption{Sensitivity of the parameters $m_{s}$, $g_{s}$, $g_{\omega}$ and $C$ as we vary $\mathrm{q_c}$(left panels), $\Lambda_M$ (middle panels) and $\Lambda_{\rho^T}$ (right panels). The three scenarios for $\kappa_{\rho}$ are also considered. The reference values are given in Tab.~\ref{tab:results_D}. The shaded region represents a 3\% deviation from the reference values.}
\label{fig:sensitivity}
\end{figure*}

\section{Results}
\label{sec:Results}

We now present our results, for which we consider our ultimate model D containing the effects of nucleon finite size and SRC, which impact mostly the short distance properties of the interaction elements. Our results are presented in the following way: We first perform a sensitivity analysis of our coupling constants and their impact on the observables against the variation of some arbitrary parameters, such as the value of $\mathrm{q}_c$ (employed for the description of the SRC) and the cut-off $\Lambda_\alpha$ and $\Lambda_{\rho^T}$ (employed to include nucleon finite size). We then perform a Bayesian exploration where the uncertainties in the quantities used in our fit, see Tab.~\ref{tab:fitting}, are explored and propagated to our predictions for nuclear matter. We also study the impact of the strength of the $\rho^T$ coupling constant on the effective masses (Dirac and Landau) and show its impact on the isoscalar and isovector components of the Landau effective mass, before finally discussing the predictions of our model for low densities as compared to other employed models.

\subsection{Sensitivity analysis}

We first test the dependence of our results on a few parameters which are commonly fixed. In Fig.~\ref{fig:sensitivity}, we analyse the impact of $\mathrm{q}_c$ (controlling the strength of the SRC) and the cut-off parameters $\Lambda_\alpha$ and $\Lambda_{\rho^T}$ (controlling the nucleon finite size) on the parameters of the model $m_{s}$, $g_{s}$, $g_{\omega}$ and $C$. Note that, in this sensitivity analysis, we systematically refit the parameters and we provide the ratio of the new parameters over reference ones, given in Tab.~\ref{tab:results_D}. In Fig.~\ref{fig:sensitivity} the values of the parameters $\mathrm{q}_c$, $\Lambda_\alpha$ and $\Lambda_{\rho^T}$ are varied by about 20\%
around the values that we have adopted in Tab.~\ref{tab:fixedparams}. We observe that $\mathrm{q}_c$ and $\Lambda_\alpha$ have a very small impact on the model parameters, less than 3\% correction. This shows that a small change in the description of the SRC does not have a big impact on the model parameters. This is also the case for the nucleon finite size probed by the meson fields, except for the $\rho^T$ case. The cut-off parameter $\Lambda_{\rho^T}$ is indeed the one which has the largest impact on the parameters since it can change them by 3-5\% at most. We can notice that the largest correlation are observed between $\Lambda_{\rho^T}$ and the scalar-field coupling constant $g_s$. Both $s$ and $\rho^T$ are attractive, and then the condition to reproduce the saturation properties implies a compensation between these two terms, which translates into an anti-correlation pattern, as observed in Fig.~\ref{fig:sensitivity}. In conclusion, we however conclude that the variation of the parameters $\mathrm{q}_c$, $\Lambda_\alpha$ and $\Lambda_{\rho^T}$ weakly impact, at most 5\%, but less than 3\% in most cases, the model parameters.

\begin{figure}[t]
\centering
\includegraphics[width=0.5\textwidth]{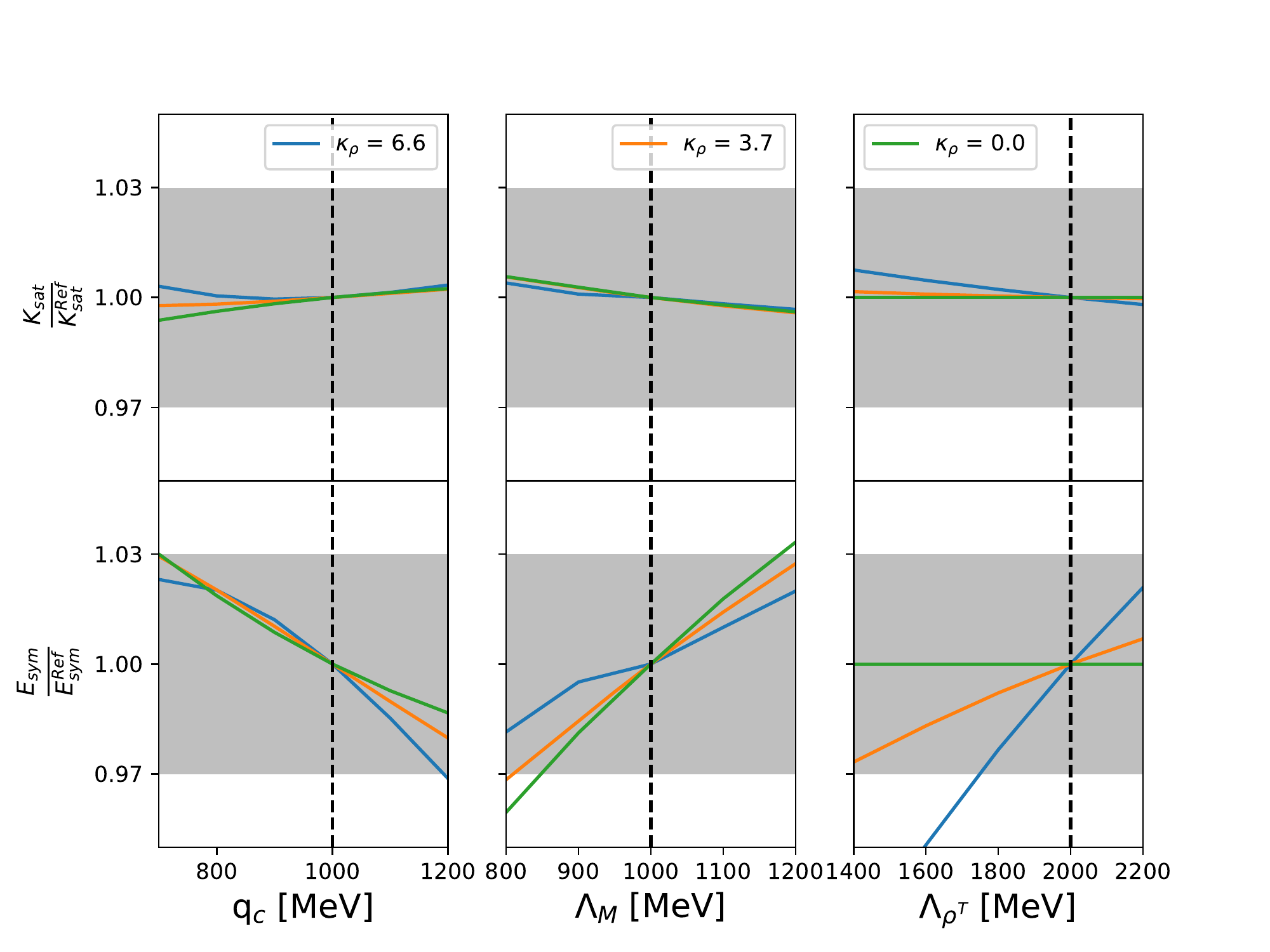}
\caption{Sensitivity of the observables $K_{\sat}$ and $E_{\sym}$ as we vary $\mathrm{q_c}$(left column), $\Lambda_M$ (middle column) and $\Lambda_{\rho^T}$ (right column). The 3 scenarios for $\kappa_{\rho}$ are also considered. The values are normalised to the the respective ones found in Tab.~\ref{tab:results_D}. The shaded region represents a 3\% deviation from the reference values.}
\label{fig:observables}
\end{figure}

The impact of the same three parameters on the incompressibility modulus $K_\sat$ and on the symmetry energy $E_\sym$ are shown in Fig.~\ref{fig:observables}. The sensibility of $K_\sat$ on these three parameters as well as on the $\rho^T$ coupling constant is small, less than 1\%, while $E_\sym$ is more impacted. The SRC can modify $E_\sym$ by 3\% at most, similarly to the finite size of the nucleon, except in the $\rho^T$ channel. As expected, the $\rho^T$ channel impacts $E_\sym$ by up to 5\%. We however conclude that globally the NEP $K_\sat$ and $E_\sym$ are weakly impacted by the parameters  $\mathrm{q}_c$, $\Lambda_\alpha$ and $\Lambda_{\rho^T}$. In conclusion, the parameters $\mathrm{q}_c$, $\Lambda_\alpha$ and $\Lambda_{\rho^T}$ can safely by fixed to arbitrary values since they will not impact substantially our results. This is the strategy we adopt in the following and we safely fix the parameters $\mathrm{q}_c$, $\Lambda_\alpha$ and $\Lambda_{\rho^T}$ to their reference value given in Tab.~\ref{tab:fixedparams}.

\subsection{Error propagation from a Bayesian analysis}

A Bayesian approach is employed to compare our model predictions, represented by a set of parameters $\{a_i\}$ with the present data~\cite{Dobaczewski:2014}. The so-called posterior probability, is the probability associated to a given model considering a set of data. It is defined as
\begin{equation}\label{e13}
P(\{a_i\} \mid \textrm{data})\sim P(\textrm{data}\mid \{a_i\})\times P(\{a_i\}),
\end{equation}
where $P(\textrm{data}\mid \{a_i\})$ is the likelihood function representing the ability of the model to reproduce a set of measurements and $P(\{a_i\})$ is the prior probability, which represents the \textsl{a-priori} knowledge on the model parameters. These model parameters are $a_2$, $a_4$, $g_{\omega}$ and $m_{s}$, which are fixed by L-QCD data and the two NEP $n_{\sat}$ and $E_{\sat}$, including their uncertainties as given in Table.~\ref{tab:fitting}. The prior distribution for the Lattice parameter $a_2$ and $a_4$ has been taken as flat for the $L1$ L-QCD analysis. 


The Bayesian framework allows us to investigate the propagation of experimental uncertainties on i) our model's parameters as well as on ii) our predictions. By marginalizing over all other parameters, one could generate a Probability Distribution Function (PDF) associated to the model parameters and to our predictions. 
We additionally couple the Bayesian framework to the Markov-Chain Monte-Carlo (MCMC) in order to guide the exploration of the uncertainties in the experimental data. More details are given in our previous study~\cite{Rahul2022}.

\begin{figure}[t]
\centering
\includegraphics[width=0.5\textwidth]{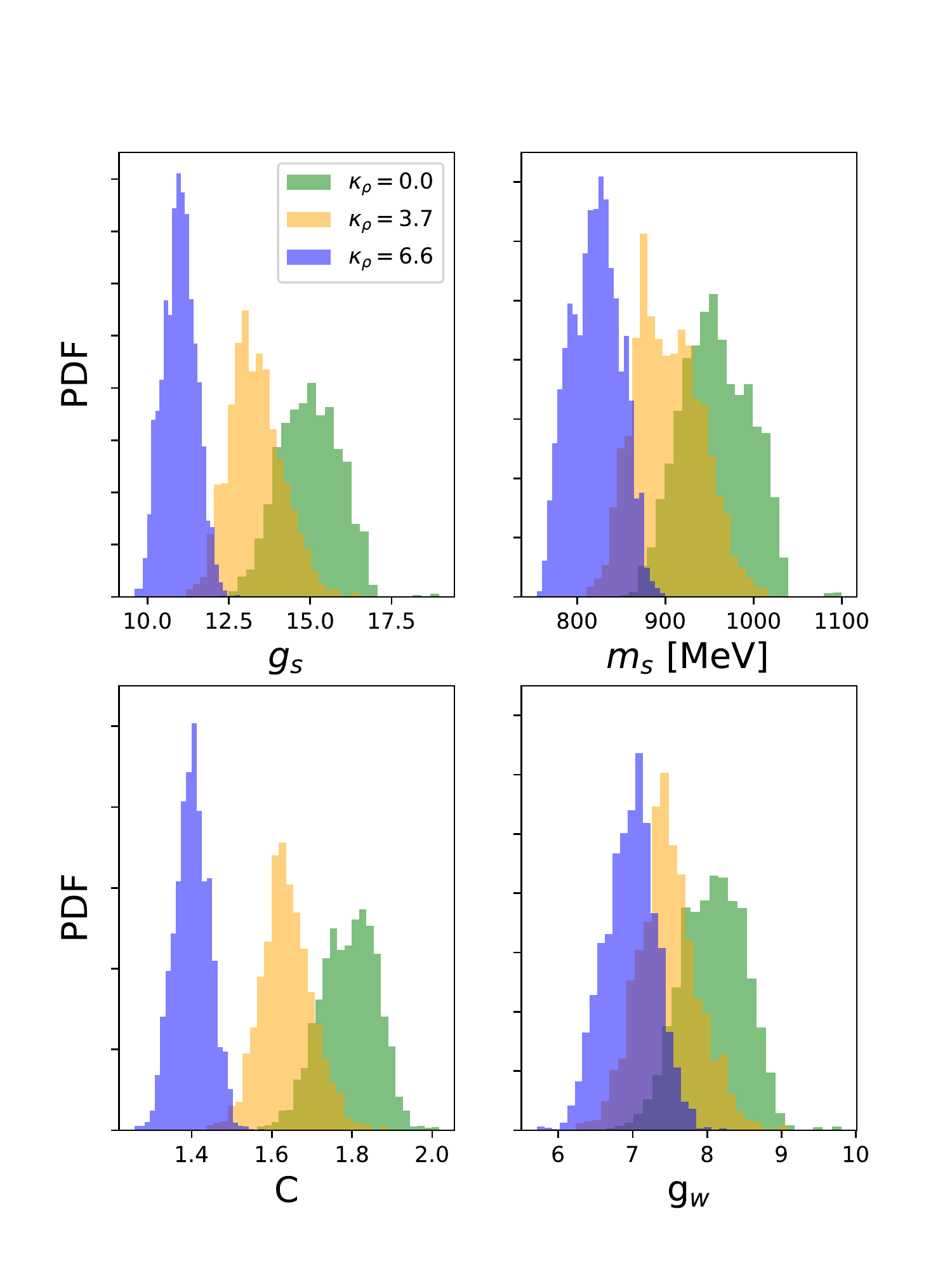}
\caption{Probability Distribution Function (PDF) for the case D with the L-QCD set L1 and the three $\kappa_\rho$ scenarios, adjusted to reproduce the saturation properties n$_{\sat}$ and E$_{\sat}$ of Tab.~\ref{tab:fitting}.}
\label{fig:couplings_LQCD-1}
\end{figure}

The PDFs obtained for the model parameters $m_{s}$, $g_{s}$, $g_{\omega}$ and $C$ are shown in Fig~\ref{fig:couplings_LQCD-1}. The peak of the PDF coincide well with the best parameter sets given in Tab.~\ref{tab:results_D}, showing a large impact of the choice of the $\rho^T$ coupling constant on the model parameters. In addition, the Bayesian analysis provides the width of the distribution in the model parameters: The widths are direct reflection of the experimental uncertainties.

\begin{figure}[t]
\centering
\includegraphics[width=0.5\textwidth]{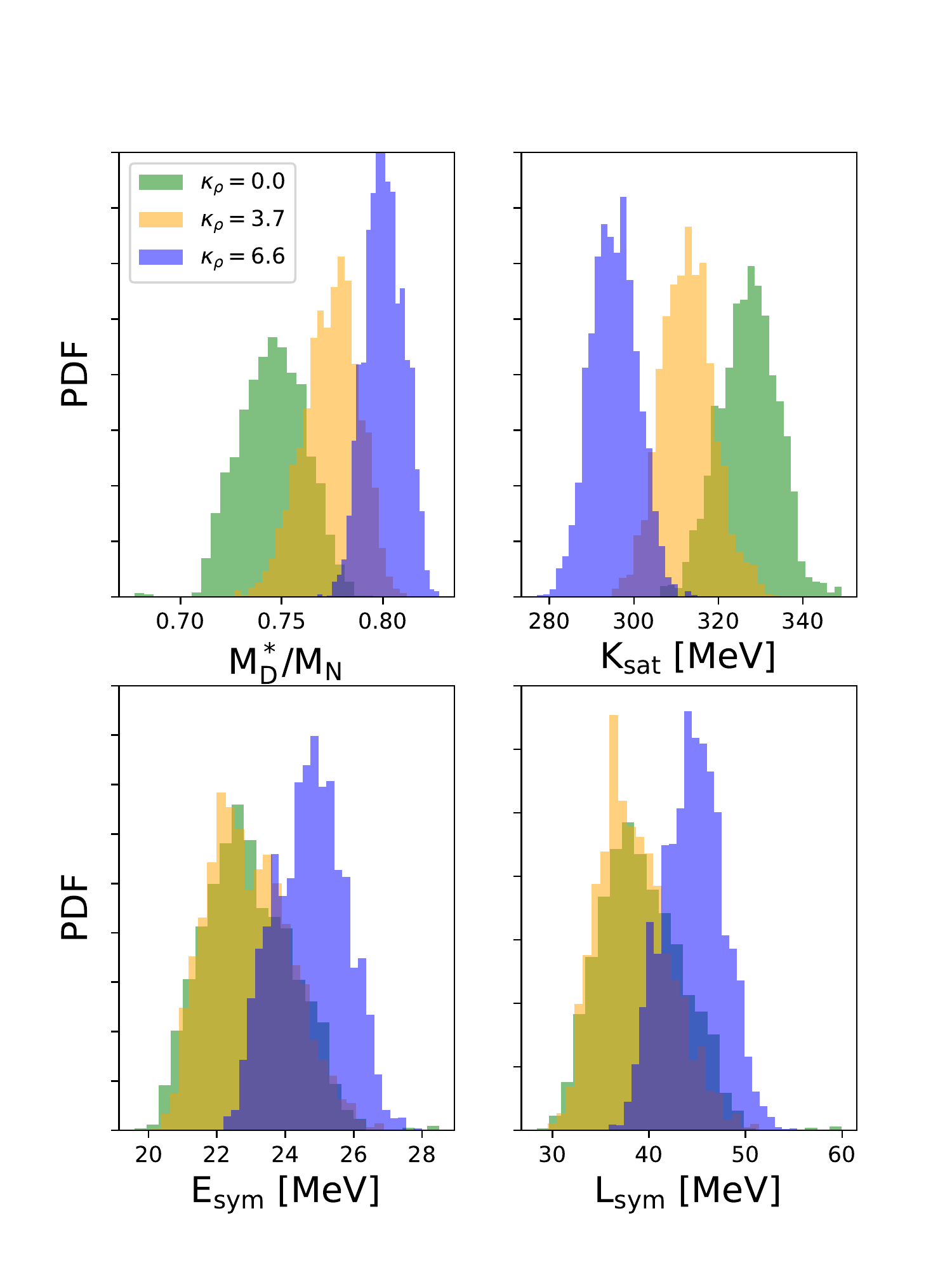}
\caption{Probability Distribution Function (PDF) for the NEP parameters M$^*_{D}$, the incompressibility modulus K$_{\sat}$, the symmetry energy E$_{\sym}$ and the slope of the symmetry energy L$_{\sym}$ for the case D with the L-QCD set L1 and the three $\kappa_\rho$ scenarios, adjusted to reproduce the saturation properties n$_{\sat}$ and E$_{\sat}$ of Tab.~\ref{tab:fitting}.}\label{fig:NEP_LQCD-1}
\end{figure}

Similarly, the Bayesian approach provides the PDF for our model predictions, see Fig.~\ref{fig:NEP_LQCD-1}. Considering $L1$ L-QCD constraints, Fig.~\ref{fig:NEP_LQCD-1} shows the PDF for the Dirac mass $M^*_{D}$, the incompressibility modulus $K_{\sat}$, the symmetry energy $E_{\sym}$ and its slope $L_\sym$. Again, there is a large influence of the $\rho^T$ coupling constant. Note however that in the iso-vector channel, $E_\sym$ and $L_\sym$ the case NRT and WRT are quite degenerate, and differences becomes marked only if large values for the $\rho^T$ are considered. The Dirac mass ranges from 0.72 up to 0.85, the incompressibility modulus from 285 to 340~MeV, with low values obtained for the SRT case. The symmetry energy spans from 21 to 27~MeV and its slope from 32 up to 50~MeV. The largest values for $E_\sym$ and $L_\sym$ are obtained in the SRT case.

\begin{table}[t]
\tabcolsep=0.32cm
\def\arraystretch{1.3}
\caption{Mean and standard deviation values obtained for the model parameters and some NEP as extracted from Figs.~\ref{fig:couplings_LQCD-1} and \ref{fig:NEP_LQCD-1}.}
\label{tab:PDF}
\begin{tabular}{c|ccc}
\hline
 & NRT & WRT & SRT \\
\hline
$g_s$ & $15.01\pm 0.93$ & $13.32\pm 0.81$ & $10.99\pm 0.50$ \\
$m_s$  & $957\pm 38$ & $903\pm 37$ & $821\pm 27$ \\
$C$  & $1.79\pm 0.07$ & $1.64\pm 0.06$ & $1.40\pm 0.04$ \\
$g_\omega$ & $8.08\pm 0.42$ & $7.47\pm 0.40$ & $6.97\pm 0.34$ \\
$K_{\sat}$ & $327\pm 7$ & $313\pm 6$ & $295\pm 6$ \\
$E_{\sym}$ & $23.0\pm 1.3$ & $23.0\pm 1.2$ & $24.7\pm 1.0$ \\
$L_{\sym}$ & $39.2\pm 4.2$ & $38.4\pm 3.7$ & $44.6\pm 3.1$ \\
\hline
\end{tabular}
\end{table}

The centroids and standard deviations associated to each of the distributions shown in Figs.~\ref{fig:couplings_LQCD-1}
and \ref{fig:NEP_LQCD-1} are reported in Tab.~\ref{tab:PDF}.

In conclusion, the knowledge of the width distribution in the NEP shows that it is impossible to reproduce the experimental values for $K_\sat$ and $E_\sym$~\cite{Margueron2018} by exploring the experimental uncertainties used in the fit of our model parameters. The inclusion of the Fock correlations together with the nuclear finite size and the SRC, improve our prediction for the values of the NEP (which are not used in the fit) compared to the case where the Fock terms are absent, see RH$_\mathrm{CC}$ and Ref.~\cite{Rahul2022}.
The best reproduction of these NEP are however obtained for the SRT case.

\subsection{Impact of lattice QCD}

In the present study, we have reported two parameter sets extracted from L-QCD, which are named L1 and L2, see Tab.~\ref{tab:modelparameterfit}. We have mostly employed L1 for reasons which will be made more clear in this section. The results based on the L2 set are given in Tab.~\ref{tab:fitting}, which should be compared to Tab.~\ref{tab:results_D} based on L1 set. We employ in these two tables model D and all other parameters are fixed to be identical. Let us note that we fit the same saturation properties ($n_\sat$ and $E_\sat$). 
Considering the L2 set, we observe that the scalar field coupling and mass are very different from the ones we obtained from the L1 set: $m_s$ and $g_s$ are much larger than the values we obtain from the L1 set. Note however that the symmetry energy spans between 31.40 and 34.01~MeV, which is in better agreement with the experimental data~\cite{Margueron2018}. There is a decrease of $K_\sat$ and $E_\sym$ as the $\rho^T$ coupling increases, and the model considering SRT is the one which is again the closer to the experimental values for $K_\sat$ and $E_\sym$.

\begin{table*}[t]
\tabcolsep=0.075cm
\def\arraystretch{2.0}
\caption{\label{tab:results_D_L2}%
The parameters for the model D, fitted to the L2 mean value of a$_2$ and a$_4$, with the 3 values of $\kappa_{\rho}$. We also show the energy contribution of the various mesons at the Hartree and Fock levels. The delta meson contribution is negligible so it's not shown.}
\begin{tabular}{lcccc|cccc|cccccccc}
\hline
\multicolumn{5}{c|}{Parameters} & \multicolumn{4}{c}{NEP}  & \multicolumn{8}{|c}{Meson contribution to the binding energy} \\
\hline
model & $m_{s}$ & $g_{s}$ & $g_{\omega}$ & $C$ & $K_{\sat}$ & $E_{\sym}$ & $M_{D}^*/M_N$ & $E_\mathrm{K}$ & $E^{s}_\mathrm{H}$  & $E^{s}_\mathrm{F}$  & $E^{\omega}_\mathrm{H}$ & $E^{\omega}_\mathrm{F}$ & $E^{\pi}_\mathrm{F}$ & $E^{\rho^V}_\mathrm{F}$ & $E^{\rho^T}_\mathrm{F}$ & $E^{\rho^{VT}}_\mathrm{F}$ \\
 & MeV & & & & MeV & MeV & & MeV & MeV & MeV & MeV & MeV & MeV & MeV & MeV & MeV \\
\hline
RHF$_\mathrm{CC}(D,L2,\mathrm{NRT})$ & 1471 & 25.316 & 9.975 & 2.68 & 361 & 34.01 & 0.64 & 16.1 & -136.7 & 21.4 & 96.6 & -15.2 & 6.8 & -5.1 & 0.0 & 0.0 \\
RHF$_\mathrm{CC}(D,L2,\mathrm{WRT})$ & 1341 & 21.044 & 8.902 & 2.38 & 330 & 31.26 & 0.70 & 17.1 & -112.8 & 16.6 & 77.0 & -12.3 & 6.9 & -4.1 & -8.4 & 3.3 \\
RHF$_\mathrm{CC}(D,L2,\mathrm{SRT})$ & 1176 & 16.192 & 8.050 & 1.97 & 302 & 31.40 & 0.75 & 18.2 & -85.1 & 11.8 & 62.9 & -10.2 & 6.9 & -3.4 & -21.9 & 4.6 \\
\hline
\end{tabular}
\end{table*}

\begin{figure}[t]
\centering
\includegraphics[width=0.5\textwidth]{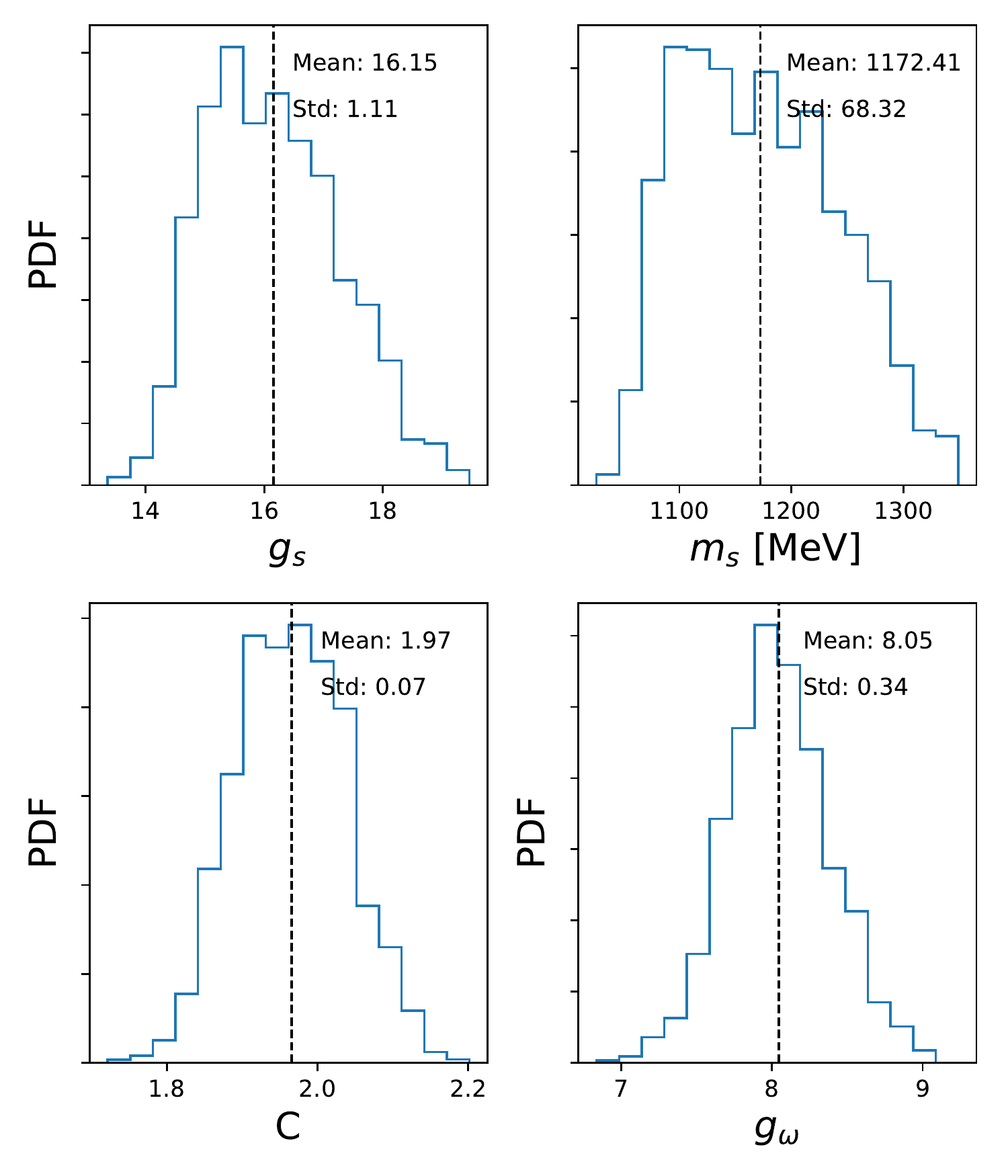}
\caption{Probability Distribution Function (PDF) for the fit parameters for the RHF$_{CC}(D,L2,SRT)$ adjusted to reproduce the saturation properties $n_{\sat}$ and $E_{\sat}$ of Tab.~\ref{tab:fitting}.}
\label{fig:couplings_LQCD-2}
\end{figure}

We perform a Bayesian study for L2 identical to the one we have performed for L1. Fig.~\ref{fig:couplings_LQCD-2} shows the results and shall be analysed in comparison to Fig.~\ref{fig:couplings_LQCD-1}. It shows that considering the uncertainties in $a_2$ and $a_4$ as given in L2 set, the distribution in the scalar field parameters $g_s$ and $m_s$ point toward values which are substantially larger than the values expected for instance from the L$\sigma$M. For this reason, we consider that our results based on L1 set are closer to the expected results. In the following, we define L1 as our reference set.

\subsection{Effective masses and nuclear empirical parameters}
\label{sec:Effective_mass}

The effective mass~\cite{Jaminon1989} is a quantity which is interesting to analyse since it determines the density of states around the Fermi level, as well as pairing properties, and low and finite-temperature properties of nuclei and nuclear matter. There are however different definitions of the effective mass in the literature, e.g. the scalar Dirac mass $M_D^*$, which emerges naturally in a covariant approach, or the Landau mass $M_L^*$ which is natural in non-relativistic approaches since it subsumes the quadratic momentum dependence of the single particle energy. These two effective masses have indeed different physical origins~\cite{Ma2004,vanDalen2005}. In terms of observables, the scalar Dirac mass $M_D^*$ contributes to the spin-orbit potential while the Landau mass $M_L^*$ influences the density of states. They are are also mainly governed by different meson couplings. 

In the following, we briefly summarize how the Landau mass is obtained in covariant approaches.
Starting from the relativistic Dirac equation,
\begin{equation}
(\bm{\gamma}\cdot\mathbf{k^*} + M^*_D)\psi = \gamma_0 E^* \psi \, ,
\end{equation}
we first derive a Schr\"odinger equivalent equation of the following form:
\begin{equation}
(\mathbf{\gamma}\cdot\mathbf{k} + M_N + U_S + \gamma_0 U_0)\psi = \gamma_0 E \psi
\end{equation}
with the new scalar and vector potentials 
\begin{equation}
U_S = \frac{\Sigma_S - M\tilde{\Sigma}_V}{1+\tilde{\Sigma}_V} \quad  \quad U_0 = \frac{\Sigma_0 + E\tilde{\Sigma}_V}{1+\tilde{\Sigma}_V}\, 
\end{equation}
where $E = E^* + \Sigma_0$ and $\tilde{\Sigma}_V = k^{-1}\Sigma_V$.
The relativistic energy-momentum relation
\begin{equation}
k^{*2} + M_D^{*2} = E^{*2} \, ,
\end{equation}
is then rewritten as
\begin{equation}
\frac{\mathrm{k}^2}{2M_N} + V_\mathrm{eq} = \frac{E^2-M_N^2}{2M_N} \, ,
\end{equation}
where $V_\mathrm{eq}$ is the equivalent Schr\"odinger potential, defined as
\begin{align}
V_\mathrm{eq}(\mathrm{k},\epsilon) =& U_S + \frac{M_N+\epsilon}{M_N}U_0 + \frac{\mathrm{k}}{2M_N}\Sigma_V \nonumber\\ & +\frac{1}{2M_N}(U_S^2 - U_0^2 + \Sigma_V^2).
\end{align}
with $E = \epsilon + M_N$.
The group velocity $v_g$ is defined as the physical velocity of the wave packet,
\begin{equation}
\label{eq:group_velocity}
v_g = \frac{d\epsilon}{d\mathrm{k}} \, .
\end{equation}
In free space, the group velocity is just $v_g = \mathrm{k}/M$. In the medium, the mass $M$ is replaced by the Landau effective mass $M_L^*$, which takes into account the presence of other nucleons in the medium. It is defined as
\begin{equation}
v_g^*=\frac{\mathrm{k}}{M^*_L} \equiv \frac{d\epsilon}{d\mathrm{k}}  
\end{equation}
Since $\epsilon$ has an explicit $\mathrm{k}$ dependence, as well as an implicit one, we obtain,
\begin{equation}
\frac{M^*_L}{M_N} = \frac{M^*_k}{M_N} . \frac{M^*_{\epsilon}}{M_N}
\end{equation}
where the so-called effective non-local mass reads,
\begin{equation}
\frac{M^*_k}{M_N} = \left[1+\frac{M_N}{\mathrm{k}}\frac{\partial}{\partial \mathrm{k} }V_\mathrm{eq}(\mathrm{k},\epsilon) \right]^{-1}
\end{equation}
and the so-called effective dynamical mass reads,
\begin{equation}
\frac{M^*_{\epsilon}}{M_N} = \left [1-\frac{\partial}{\partial \epsilon }V_\mathrm{eq}(\mathrm{k},\epsilon) \right] \, .
\end{equation}
Note that the Landau mass is defined at the Fermi surface for the neutrons and the protons.

\begin{figure}[t]
\centering
\includegraphics[width=0.5\textwidth]{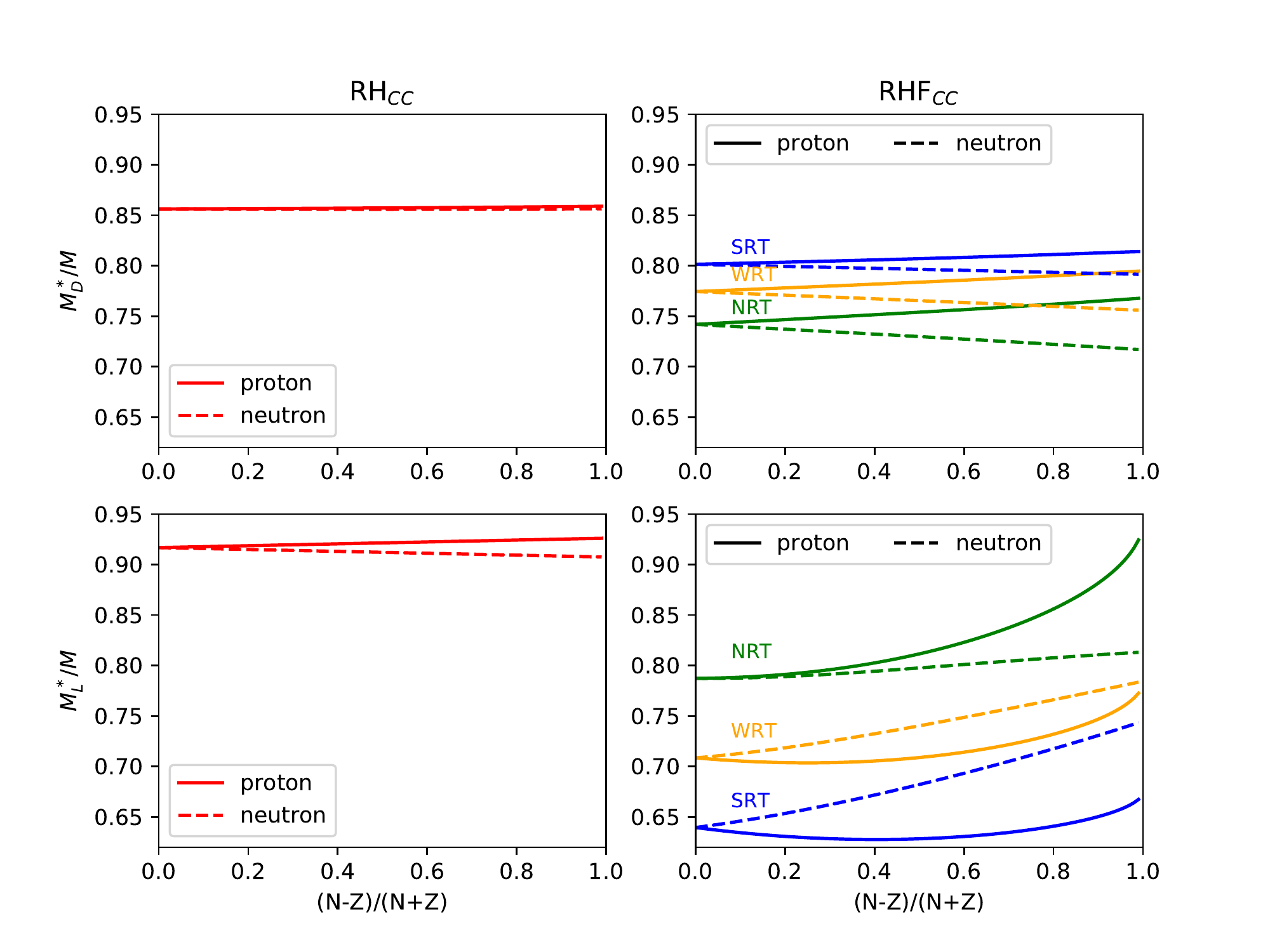}
\caption{The Landau and Dirac masses at the Fermi surface as a function of the asymmetry parameter for both proton (solid lines) and neutrons (dashed lines) in both RH$_\mathrm{CC}$ and RHF$_\mathrm{CC}$. For the RHF$_\mathrm{CC}$, the three $\rho^T$ couplings are shown.}
\label{fig:effective_mass}
\end{figure}

In Fig.~\ref{fig:effective_mass}, we show the splitting between the Landau (bottom panels) and Dirac (top panels) masses as a function of the isospin asymmetry parameter $(N-Z)/(N+Z)$. On the left panels, we show results at the Hartree approximation (RH$_\mathrm{CC}$) while on the right ones we show results at the HF approximation (model D). 

Let's start the discussion at the Hartree approximation, since it is a case where the connection with the meson field properties is the more direct. The effective masses are defined as
\begin{eqnarray}
M^*_{D,N} &=& M_N + \Sigma^D_{S,N} \,, \\
M^*_{L,N} &=& M_N - \Sigma^D_{0,N} \,.
\end{eqnarray}
where $\Sigma^D_{S,N}$ and $\Sigma^D_{0,N}$ are given by Eqs.~\eqref{eq:sigmaS_direct} and \eqref{eq:sigma0_direct}. The splittings of the effective masses are induced by different mesons: The $\delta$ meson creates the splitting of the Dirac mass, and having a very small coupling constant, this explains the small splitting observed in Fig.~\ref{fig:effective_mass} for the Dirac mass considering RH$_{CC}$, whereas it is the $\rho$ meson which create a splitting of the Landau mass, explaining the larger splitting observed in Fig.~\ref{fig:effective_mass}.

The contribution of the Fock terms can be observed by comparing left and right panels in Fig.~\ref{fig:effective_mass}: They contribute to decrease further the effective masses.
In addition, the $\rho^T$ coupling scenario plays an important role for the two effective masses: In symmetric matter, it contributes to increase the Dirac mass $M_D^*$, while it decreases the Landau mass $M_L^*$.
For asymmetric matter, the $\rho^T$ coupling scenario also plays a role in the strength of the splittings. It reduces the splitting of the Dirac mass in asymmetric matter, and for the Landau mass its effect is even more spectacular since it changes the sign of the splitting: without $\rho^T$ coupling (NRT) we obtain $M_L^*(n)-M_L^*(p)\approx -0.15$, while in the SRT case, we obtain $M_L^*(n)-M_L^*(p)\approx 0.05$-$0.1$ in neutron matter. We then conclude that the sign of the splitting of the Landau mass in neutron matter is strongly correlated with the value of the $\rho^T$ coupling.

This sign was predicted to be positive in Ref.~\cite{vanDalen2005}, based on the Bonn potential for which this coupling is large (SRT). 
According to our results, a small value, or a value compatible with the WRT case may largely reduce the splitting, or even lead to an opposite sign. It is however difficult to extract from experimental data the splitting of the effective mass. We therefore believe that the sign of the splitting of the Landau mass in neutron matter is not yet a settled question.

\begin{table*}[t]
\tabcolsep=0.15cm
\def\arraystretch{2.0}
\caption{Predictions for the Dirac and Landau masses and other NEP from the model D and different prescription for the $\rho^T$ coupling. The results of the two sets of L-QCD are shown.}
\label{tab:NEP2}
\begin{tabular}{lcccc|cccccccccccc}
\hline
model & $m_s$ & $g_{s}$ & $g_{\omega}$ & $C$ & $K_{\sat}$ & $E_{\sym}$ & $M_{D}^*/M_N$ & $M_L^*/M_N$ & $\Delta M_L^*/M_N$ & $L_\sym$ & $K_\sym$ & $Q_\sat$ \\
 & MeV & & & & MeV & MeV & &  &  & MeV & MeV & MeV \\
\hline
RHF$_\mathrm{CC}(D,L1,\mathrm{NRT})$ & 965 & 15.200 & 8.176 & 1.81 & 328 & 23.2 & 0.74 & 0.79 & -0.12 & 40.4 & -34.4 & -262  \\
RHF$_\mathrm{CC}(D,L1,\mathrm{WRT})$ & 903 & 13.296 & 7.457 & 1.64 & 314 & 22.9 & 0.77 & 0.71 & 0.01 & 38.8 & -48.7 & -332 \\
RHF$_\mathrm{CC}(D,L1,\mathrm{SRT})$ & 819 & 10.943 & 6.932 & 1.40 & 296 & 24.7 & 0.80 & 0.64 & 0.07 & 45.0 & -56.8 & -401 \\
RHF$_\mathrm{CC}(D,L2,\mathrm{NRT})$ & 1471 & 25.316 & 9.975 & 2.68 & 361 & 34.0 & 0.64 & 0.71 & -0.14 & 80.8 & -11.4 & 116\\
RHF$_\mathrm{CC}(D,L2,\mathrm{WRT})$ & 1341 & 21.044 & 8.902 & 2.38 & 330 & 31.3 & 0.70 & 0.63 & 0.01 & 66.0 & -59.1 & -113\\
RHF$_\mathrm{CC}(D,L2,\mathrm{SRT})$ & 1176 & 16.192 & 8.050 & 1.97 & 302 & 31.4 & 0.75 & 0.57 & 0.08 & 65.1 & -78.3 & -262 \\
\hline
\end{tabular}
\end{table*}

We now summarize the numerical values for the NEP and effective masses for various models exploring the prescriptions for $\rho^T$ and the L-QCD sets. Our results are given in Tab.~\ref{tab:NEP2} for the best parameter sets. Some of the results given in this table have already been given in Tabs~\ref{tab:results_D} and \ref{tab:results_D_L2}, but we decided to assemble them in a unique table in order to compare them. We find a large impact of the L-QCD set on the symmetry energy properties: the set $L2$ leads to values of $E_\sym$ in close agreement with the empirical expectation of $32\pm 2$~MeV~\cite{Margueron2018}, and values for its slope $L_\sym$ in the high domain of the expectation $60\pm 15$~MeV~\cite{Margueron2018}. The set $L1$ however lead to a smaller value of $E_\sym$ and $L_\sym$. The isoscalar NEP $Q_\sat$ is larger for the set $L2$ compared to $L1$ and the Dirac and Landau masses are larger for $L1$ compared to $L2$, but none of these values could be definitively excluded. The main drawback of the set $L2$ can be find in the large values for the scalar mass $m_s$ and coupling constant $g_s$, as we already observed in a previous discussion. For this reason, we still prefer the set $L1$ and we note that it predicts values for the incompressibility $K_\sat$ a bit larger than the expected one and values for the NEP $E_\sym$ about 8~MeV below the expected one. We show in the following that the low value of $E_{\sym}$ obtained in our best approach impacts the neutron matter EoS around saturation density.

\subsection{Prediction for low density neutron matter}

\begin{figure*}[t]
\centering
\includegraphics[width=0.9\textwidth]{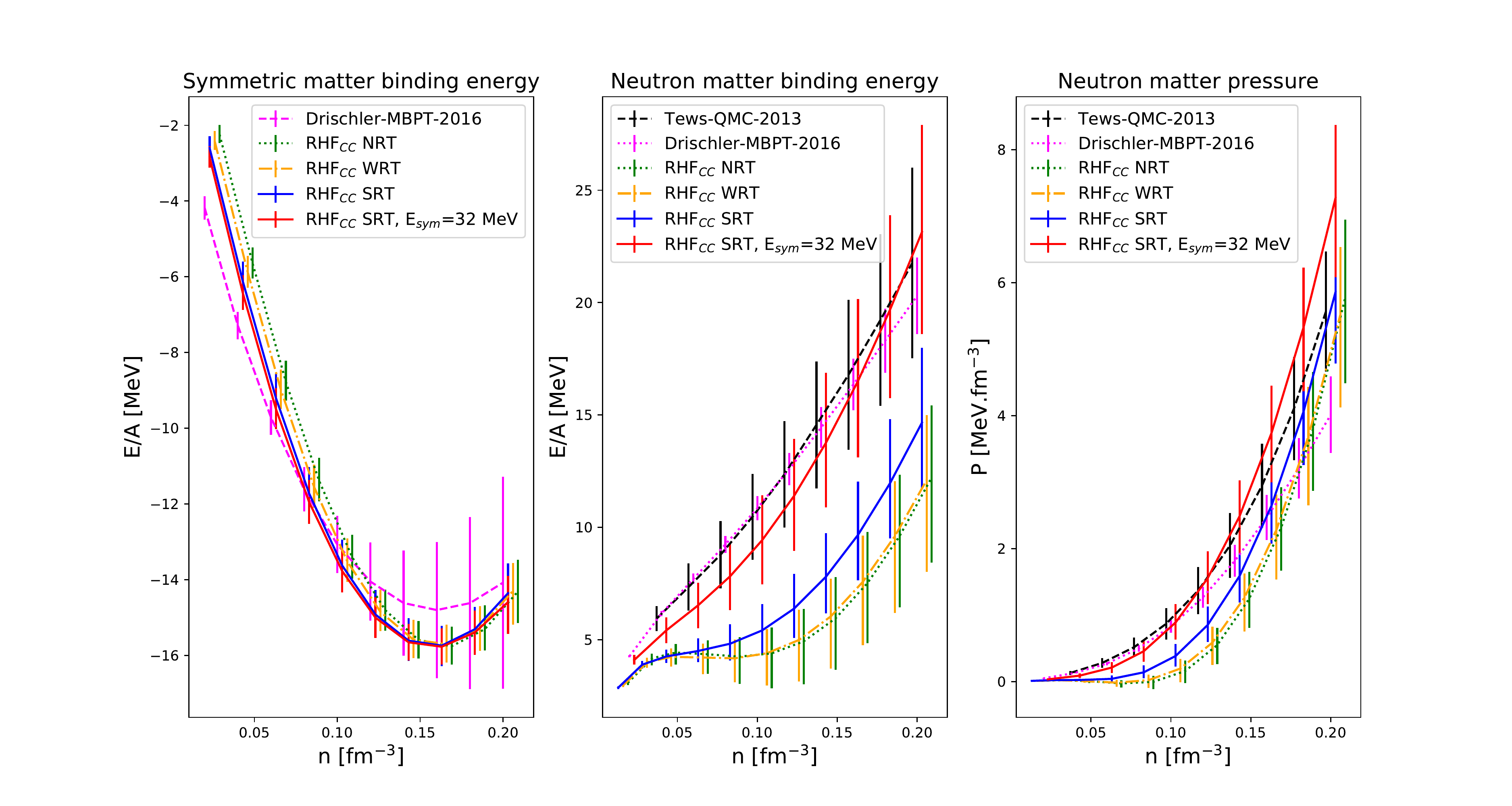}
\caption{Low density equations of state for symmetric matter's binding energy (left panel), neutron matter's binding energy (middle panel) and neutron matter's pressure(right panel). We show results from other models (see Refs.~\cite{Drischler2016,Tews2013}), and for the RHF$_{CC}$, for which we show the three $\rho^T$ cases. In our last model, we replace the constraint imposed by the quark model by the constraint to reproduce the experimental symmetry energy $E_\sym=32$~MeV and we only show the SRT.
}
\label{fig:lowden_EoS}
\end{figure*}

So far, we have generated RHF$_\mathrm{CC}$ models reproducing some saturation properties of symmetric nuclear matter ($E_\sat$ and $n_\sat$). We now study in more details the prediction of our models for neutron matter at low density, where they are compared to predictions based on $\chi$-EFT interaction and different many-body treatments.
In Fig.~\ref{fig:lowden_EoS} are shown the binding energy in symmetric and neutron matter, as well as the pressure in neutron matter. The uncertainties in our models correspond to 95\% confidence interval from our Bayesian samples. We compare our results to other calculations: Drischler' 2016 many-body perturbation theory based on $\chi$-EFT two and three-body interactions in symmetric and neutron matter~\cite{Drischler2016}. In neutron matter, we additionally compare our results to the predictions by Tews~\cite{Tews2013} from QMC approach.

We remind that the symmetry energy we find for the case D is too low compared to the experimental one, and since the saturation energy $E_{\sat}$ is fixed, we expect our prediction for the binding energy in neutron matter to be lower. This is indeed what we observe in the middle panel. By looking at the pressure, we see that EoS is soft at the beginning and then it starts getting stiffer, overcoming the other EoS at $\sim$ 0.16~fm$^{-3}$. We also note that for the NRT and WRT cases, the pressure is slightly negative at the beginning, indicating an instability in the model.

Finally, we show the prediction of our RHF$_\mathrm{CC}$ model in the case for which the value symmetry energy is used in the fit of the model parameters to get the $\rho$ vector coupling constant $g_\rho$, instead of the value fixed by the quark model (only the SRT scenario is shown, see the next subsection for more information). We observe a much better agreement with the chiral prediction for the neutron matter EoS, as we discuss hereafter.

\subsection{Departing from the quark model constraint for $g_\rho$}
\label{sec:quark model}

We remind that the coupling constant $g_\rho$ is obtained from the quark model, giving $g_\rho = g_\omega /3$. This constrain however may be removed and replaced by another one, e.g. fixing $g_\rho$ to get the experimental value for $E_\sym$, in order to estimate which value the $\rho$ coupling should get in order that our results reproduce correctly the energy of neutron matter. For that we take the mean values of the L-QCD parameters from Tab.~\ref{tab:fitting}, and we now add $E_{\sym} =  32 ~\pm~ 2 $ MeV to be satisfied by the fitting procedure, instead of the value imposed by the quark model. The results obtained are shown in Tab.~\ref{tab:noquarkmodel_results}.

The symmetry energy is governed by the isovector channel, so the $\rho$ interaction, and since our symmetry energy was lower than the empirical value, we expect $g_\rho$ to increase to make up for this. However, in order to maintain the energy saturation which was mainly governed by the scalar and vector fields $s$ and $\omega$, the attractive $\rho$ channel should be compensated by decreasing their roles, and this is indeed what we observe in the table. The decrease in $g_s$ is also accompanied by a decrease in $m_s$ to maintain a constant $a_2$ (see Eq.~\eqref{eq:LQCD}).

\begin{table}[t]
\tabcolsep=0.12cm
\def\arraystretch{1.3}
\caption{Comparison of the predictions for the model parameters and some NEP imposing the quark model constraint $g_\rho=g_\omega/3$ (same as in Tab.~\ref{tab:NEP2}) or imposing the symmetry energy $E_\sym=32$~MeV. The quantities $m_s$, $K_\sat$ and $E_\sym$ are given in MeV, the other quantities are dimensionless. Model D and L1 set for L-QCD are considered.}
\label{tab:noquarkmodel_results}
\begin{tabular}{c|ccc|ccc}
\hline
 & \multicolumn{3}{c|}{Imposing $g_\rho=g_\omega/3$} & \multicolumn{3}{c}{Imposing $E_\sym=32$~MeV} \\
 & NRT & WRT & SRT & NRT & WRT & SRT \\
\hline
$g_s$ & 15.200 & 13.296 & 10.943 & 14.131 & 10.381 & 7.732   \\
$m_s$  & 965 & 903 & 819 & 931 & 798 & 688  \\
$g_\omega$ & 8.176 & 7.457 & 6.932 & 7.951 & 6.497 & 6.091 \\
$g_\rho$  & 2.725 & 2.486 & 2.311 & 3.957 & 3.613 & 3.014 \\
$g_\omega/g_\rho$ & 3 & 3 & 3 & 2.00 & 1.80 & 2.02 \\
$C$ & 1.81 & 1.64 & 1.40 & 1.71 & 1.34 & 1.04 \\
$K_{\sat}$ & 328 & 314 & 296 & 322 & 292 & 272 \\
$E_{\sym}$ & 23.2 & 22.9 & 24.7 & 32 & 32 & 32 \\
$g^\prime$ & 0.08 & 0.23 & 0.48 & 0.08 & 0.39 & 0.76 \\
\hline
\end{tabular}
\end{table}

A more detailed comparison of the prediction based on the quark model (fixing $g_\rho=g_\omega/3$) or on the direct fit to the symmetry energy (imposing $E_\sym=32$~MeV) is shown in Tab.~\ref{tab:noquarkmodel_results}. It is interesting to note that the value of the incompressibility modulus $K_{\sat}$ is also decreasing when we impose the fit to $E_\sym$. In the SRT, the value we obtain, $K_\sat= 272 ~\pm~ 8 $ MeV is even close to compatible with the expectation from nuclear experiments. The values of the coupling constant $g_\rho$ required to better reproduce the symmetry energy are about $20$-$25$\% larger than the one imposed by the quark model. It is therefore not a strong violation of the quark model, and such a deviation 
could be understood, for instance, from a modeling of the decrease of the $\rho$ mass by about 20\% at saturation density, with respect to its bare mass in vacuum.

In Tab.~\ref{tab:noquarkmodel_results}, the value of the spin-isospin Landau-Migdal parameter $g^\prime$ is shown in the two scenarios for the $g_\rho$ coupling constant: it increases in the case when $E_\sym$ is fitted, but remain compatible with the expectation value of about 0.6 that we have discussed previously.

Finally, the prediction of this new model for the neutron matter EoS is shown in Fig.~\ref{fig:lowden_EoS}. Since we get a better agreement with nuclear physics in the iso-vector channel, it is not surprising that the neutron matter EoS is better reproduced with the new model. Indeed we observe that the new EoS is closer to the $\chi$-EFT ones within the uncertainties bars.

\section{Conclusions}
\label{sec:conclusion}

In this paper, we have investigated the effect of the chiral potential and of the nucleon polarisability in the relativistic Hartree-Fock framework in continuation of our previous study based on the relativistic Hartree approximation~\cite{Rahul2022}. The model parameters are fitted from Lattice QCD, the quark model and two empirical parameters ($n_\sat$ and $E_\sat$) are used. We have additionally included the nucleon finite size effects with form factors and the SRC with the Jastrow ansatz, inducing additional parameters in the model and we have checked that our results are only weakly influenced by these additional parameters. They are then fixed to some arbitrary values. 

The tensor component of the $\rho$ meson is varied from the NRT ($\kappa_\rho=0$) to the SRT ($\kappa_\rho=6.6$) case. We have shown that the tensor contribution to the $\rho$ meson is contributing quite substantially to several observables, such as for instance the symmetry energy $E_{\sym}$, the incompressibility modulus $K_{\sat}$, the Dirac and Landau masses $M^*_D$ and $M^*_L$, etc. We have employed two empirical parameters, $K_\sat$ and $E_\sym$, not used in the fit, to evaluate the goodness of our models. It comes that the value $\kappa_\rho=6.6$ (SRT) is the one which allows these two quantities to be the closest to their empirical values. From our analysis, we find that in general the SRT scenario is the most plausible one.

We have investigated the question of the splitting of the Landau mass $M^*_L$ in neutron matter between neutrons and protons. We found that the strength and the sign of this splitting is largely given by the size of the $\rho^T$ coupling. The value suggested in Ref.~\cite{vanDalen2005} (a splitting of about 0.1) is compatible with the SRT scenario, but a weaker $\rho^T$ coupling would lead to a smaller splitting, which could also change its sign for very small value of the $\rho^T$ coupling. Since the value of the $\rho^T$ coupling is not clearly fixed, we conclude that the splitting of the effective Landau mass may not be a settled question.

We have also replaced the quark model constraint on $g_\rho$ by the requirement to reproduce the symmetry energy $E_\sym$. The new EoS is in better agreement not only for the symmetry energy properties and the neutron matter predictions, but also for the incompressibility modulus (in the SRT scenario). The spin isospin Landau-Migdal parameter $g^\prime$ is also in good agreement with experimental expectations. In conclusion, we have obtained a RHF$_\mathrm{CC}$ model which is in a good agreement with the expectation from nuclear physics around saturation density. This is therefore a model which is satisfactory enough to analyse its predictions for neutron star matter. Our next step will therefore be to perform studies in this direction.

\begin{acknowledgements}
We would like to acknowledge Wenhui Long for sharing his RHF code in nuclear matter, from which the present models has been implemented. M.C., J.M., H.H. and G.C. are supported by the CNRS IN2P3 MAC project, and benefit from PHAROS COST Action MP16214.
R.S. is supported by the U.S. Department of Energy, Office of Science, Office of Nuclear Physics, through the Los Alamos National Laboratory. Los Alamos National Laboratory is operated by Triad National Security, LLC, for the National Nuclear Security Administration of U.S. Department of Energy (Contract No. 89233218CNA000001).
R.S. also receives support from the Laboratory Directed Research and Development program of Los Alamos National Laboratory under project number 20220541ECR.
R.S. also acknowledges support from the Nuclear Physics from Multi-Messenger Mergers (NP3M) Focused Research Hub which is funded by the National Science Foundation under Grant Number 21-16686.
This work is supported by the STRONG-2020 network from the European Union's Horizon 2020 research and innovation program under grant agreement No. 824093 (H.H.); the LABEX Lyon Institute of Origins (ANR-10-LABX-0066) of the \textsl{Universit\'e de Lyon} for its financial support within the program \textsl{Investissements d'Avenir} (ANR-11-IDEX-0007) of the French government operated by the National Research Agency (ANR).
\end{acknowledgements}


\appendix

\section{Explicit expressions for the direct contributions to the energy and to the self-energies}
\label{app:direct}

At the Hartree level, we have the direct terms contributions:
\begin{align}
\Sigma^D_{S,N} &= -\frac{g^2_{s}}{m^2_{s}}n_s - \frac{g^2_{\delta}}{m^2_{\delta}}(n_{sN}-n_{s\bar{N} } ), \\ 
\label{eq:sigmaS_direct}
\Sigma^D_{0,N} &= \frac{g^2_{\omega}}{m^2_{\omega}}n + \frac{g^2_{\rho}}{m^2_{\rho}}(n_{sN} - n_{s\bar{N} } ), \\
\label{eq:sigma0_direct}
\Sigma^D_{V,N} &= 0
\end{align}

In what follows, we will write the exchange part of the self-energies.

\section{Explicit expressions for the exchange self-energies in the case A (Pure HF calculation)}
\label{app:selfA}

\subsection{\underline{$s$ coupling}}

\setlength{\belowdisplayskip}{1pt}

\begin{subequations}
\begin{align}
\Sigma^{E,s}_{S,N} &= \frac{1}{4}g_{s}^{*^2} \int{\frac{d {\bf p'}}{(2\pi)^3}D_{s}(\mathrm{q})2\hat{M_N}(\mathrm{p'})} + \Sigma_S^{rg}, \\
\Sigma^{E,s}_{0,N} &= \frac{1}{4}g_{s}^{*^2} \int{\frac{d {\bf p'}}{(2\pi)^3}D_{s}(\mathrm{q})2f_N(\mathrm{p'})}, \\
\Sigma^{E,s}_{V,N} &= -\frac{1}{4}g_{s}^{*^2} \int{\frac{d {\bf p'}}{(2\pi)^3}D_{s}(\mathrm{q})2\tilde{\bf p} \cdot \hat{P_N}(\mathbf{p}^\prime)}, 
\end{align}
The rearrangement term $\Sigma_S^{rg}$ gives  momentum independent contributions to the scalar self energy, one due to the dependence of the effective coupling constant on the scalar field, i.e $g^*_s(\bar{s})$~:
\begin{equation}
\label{eq:rg1}
\Sigma_{S\,(rg\,1)}=-\,2\,\frac{\tilde{\kappa}_{NS}}{m^{*2}_{s}}\,\epsilon_{Fock}^{(s)}
\end{equation}
with $\tilde{\kappa}_{NS}$ defined in Eq.~\eqref{eq:kappa_tilde},
and the other contribution due to the dependence of the effective scalar mass $m^*_s(\bar{s})$ on the scalar field~:
\begin{eqnarray}
\label{eq:rg2}
\Sigma_{S\,(rg\,2)}&=&	\frac{\partial m^{*2}_{s}}{\partial n_s}\,\frac{g_s^{*2}}{2}\, \int \frac{d{\bf p}}
{(2\pi)^3}\frac{d{\bf p}'}{(2\pi)^3}
\frac{\partial D_s(\mathrm{q})}{\partial m^*_s(\bar{s})}\nonumber\\
& &\sum_{N}\left(1+\hat{M}(\mathrm{p})\hat{M}(\mathrm{p'})-\hat{P}(\mathbf{p})\cdot\hat{P}(\mathbf{p}^\prime)\right)_N. \nonumber \\
\end{eqnarray}
where the derivative of the in-medium sigma mass with respect to the scalar density is~:
\begin{eqnarray}
\frac{\partial m^{*2}_{s}}{\partial n_s}&=&	\left(v'''(\bar s)\,+\,\frac{\partial\tilde{\kappa}_{NS}}{\partial\bar s}_,n_s\right)\,
\frac{\partial\bar s}{\partial n_s}\,+\,\tilde{\kappa}_{NS}\nonumber\\
&=&
\tilde{\kappa}_{NS}\,-\,\frac{g^{*}_{s}}{m^{*2}_{s}}\left(v'''(\bar s)\,+\,\frac{\partial\tilde{\kappa}_{NS}}{\partial\bar s}
\,n_s\right).
\end{eqnarray}
\end{subequations}

\subsection{\underline{$\delta$ coupling}}

\begin{subequations}
\begin{align}
\Sigma^{E,\delta}_{S,N} &= \frac{1}{4}g_{\delta}^2 \int{\frac{d {\bf p'}}{(2\pi)^3}D_{\delta}(\mathrm{q})2\hat{M_N}(\mathrm{p'})} \nonumber \\ 
& \hspace{1.0cm} +2\times \left(N\longmapsto\bar{N}\right), \\
\Sigma^{E,\delta}_{0,N} &= \frac{1}{4}g_{\delta}^2 \int{\frac{d {\bf p'}}{(2\pi)^3}D_{\delta}(\mathrm{q})2f_N(\mathrm{p'})} \nonumber \\ 
& \hspace{1.0cm} +2\times \left(N\longmapsto\bar{N}\right), \\
\Sigma^{E,\delta}_{V,N} &= -\frac{1}{4}g_{\delta}^2 \int{\frac{d {\bf p'}}{(2\pi)^3}D_{\delta}(\mathrm{q})2\tilde{\bf p} \cdot \hat{P_N}(\mathbf{p}^\prime)} \nonumber \\
& \hspace{1.0cm} +2\times \left(N\longmapsto\bar{N}\right).
\end{align}
\end{subequations}
\subsection{\underline{$\omega$ coupling}}

\begin{subequations}
\begin{align}
\Sigma^{E,\omega}_{S,N} &= -\frac{1}{4}g_{\omega}^2 \int{\frac{d {\bf p'}}{(2\pi)^3}D_{\omega}(\mathrm{q})8\hat{M_N}(\mathrm{p'})}, \\
\Sigma^{E,\omega}_{0,N} &= \frac{1}{4}g_{\omega}^2 \int{\frac{d {\bf p'}}{(2\pi)^3}D_{\omega}(\mathrm{q})4f_N(\mathrm{p'}) }, \\
\Sigma^{E,\omega}_{V,N} &= -\frac{1}{4}g_{\omega}^2 \int{\frac{d {\bf p'}}{(2\pi)^3}D_{\omega}(\mathrm{q})4\tilde{\bf p} \cdot \hat{P_N}(\mathbf{p}^\prime) }, 
\end{align}
\end{subequations}

\subsection{\underline{$\rho^V$ coupling}}

\begin{subequations}
\begin{align}
\Sigma^{E,\rho}_{S,N} &= -\frac{1}{4}g_{\rho}^2 \int{\frac{d {\bf p'}}{(2\pi)^3}D_{\rho}(\mathrm{q})~8\hat{M_N}(\mathrm{p'})} \nonumber \\ 
& \hspace{1.0cm} +2\times \left(N\longmapsto\bar{N}\right), \\
\Sigma^{E,\rho}_{0,N} &= \frac{1}{4}g_{\rho}^2 \int{\frac{d {\bf p'}}{(2\pi)^3}D_{\rho}(\mathrm{q})~4f_N(\mathrm{p'})} \nonumber \\
& \hspace{1.0cm} +2\times \left(N\longmapsto\bar{N}\right), \\
\Sigma^{E,\rho}_{V,N} &= -\frac{1}{4}g_{\rho}^2 \int{\frac{d {\bf p'}}{(2\pi)^3}D_{\rho}(\mathrm{q})~4\tilde{\bf p} \cdot \hat{P_N}(\mathbf{p}^\prime) } \nonumber \\
& \hspace{1.0cm} +2\times \left(N\longmapsto\bar{N}\right).
\end{align}
\end{subequations}

\subsection{\underline{$\rho^{VT}$ coupling}}

\begin{subequations}
\begin{align}
\Sigma^{E,\rho^{VT}}_{S,N} &= -\frac{1}{4} \left (\frac{g_{\rho}f_{\rho}}{2M_N} \right ) \int{}\frac{d {\bf p'}}{(2\pi)^3}D_{\rho}(\mathrm{q})~12{\bf q} \cdot \hat{P_N}(\mathbf{p}^\prime)) \nonumber \\
& \hspace{1.0cm} +2\times \left(N\longmapsto\bar{N}\right), \\
\Sigma^{E,\rho^{VT}}_{0,N} &= 0 \\
\Sigma^{E,\rho^{VT}}_{V,N} &= -\frac{1}{4}\left (\frac{g_{\rho}f_{\rho}}{2M_N} \right ) \int{\frac{d {\bf p'}}{(2\pi)^3}D_{\rho}(\mathrm{q})~12\hat{M_N}(\mathrm{p'})} \nonumber \\
& \hspace{1.0cm} +2\times \left(N\longmapsto\bar{N}\right).
\end{align}
\end{subequations}

\subsection{\underline{$\rho^{T}$ coupling}}

\begin{subequations}
\begin{align}
\Sigma^{E,\rho^T}_{S,N} &= \frac{1}{4}\left(\frac{f_{\rho}}{2M}\right)^2 \int{\frac{d {\bf p'}}{(2\pi)^3}\mathrm{q}^2D_{\rho}(\mathrm{q})~6\hat{M_N}(\mathrm{p'})} \nonumber \\
& \hspace{1.0cm} +2\times \left(N\longmapsto\bar{N}\right), \\
\Sigma^{E,\rho^T}_{0,N} &= \frac{1}{4}\left(\frac{f_{\rho}}{2M}\right)^2 \int{\frac{d {\bf p'}}{(2\pi)^3}\mathrm{q}^2D_{\rho}(\mathrm{q})~2f_N(\mathrm{p'})} \nonumber \\
& \hspace{1.0cm} +2\times \left(N\longmapsto\bar{N}\right), \\
\Sigma^{E,\rho^T}_{V,N} &= -\frac{1}{4}\left(\frac{f_{\rho}}{2M}\right)^2 \int{\frac{d {\bf p'}}{(2\pi)^3}}D_{\rho}(\mathrm{q}) \nonumber \\
&\times \left[ 2\mathrm{q}^2(\tilde{\bf p} \cdot \hat{P_N}(\mathbf{p}^\prime)) -8(\tilde{\bf p}\cdot{\bf q}) \left(\hat{P_N}(\mathbf{p}^\prime) \cdot{\bf q} \right ) \right ] \nonumber \\
& +2\times \left(N\longmapsto\bar{N}\right).
\end{align}
\end{subequations}

\subsection{\underline{$\pi$ coupling}}

\begin{subequations}
\begin{align}
\Sigma^{E,\pi}_{S,N} &= \frac{1}{4}\left(\frac{g_A}{2f_\pi}\right)^2 \int{\frac{d {\bf p'}}{(2\pi)^3}\mathrm{q}^2D_{\pi}(\mathrm{q})~2\hat{M_N}(\mathrm{p'})} \nonumber \\
& \hspace{1.0cm} +2\times \left(N\longmapsto\bar{N}\right), \\
\Sigma^{E,\pi}_{0,N} &= \frac{1}{4}\left(\frac{g_A}{2f_\pi}\right)^2 \int{\frac{d {\bf p'}}{(2\pi)^3}\mathrm{q}^2D_{\pi}(\mathrm{q})~2f_N(\mathrm{p'})} \nonumber \\
& \hspace{1.0cm} +2\times \left(N\longmapsto\bar{N}\right), \\
\Sigma^{E,\pi}_{V,N} &= -\frac{1}{4}\left(\frac{g_A}{2f_\pi}\right)^2 \int{\frac{d {\bf p'}}{(2\pi)^3}}D_{\pi}(\mathrm{q}) \nonumber \\
& \times \left[ 2\mathrm{q}^2(\tilde{\bf p} \cdot \hat{P_N}(\mathbf{p}^\prime)) -4(\tilde{\bf p}\cdot{\bf q} ) \left(\hat{P_N}(\mathbf{p}^\prime) \cdot{\bf q} \right ) \right ] \nonumber \\
& +2\times \left(N\longmapsto\bar{N}\right).
\end{align}
\end{subequations}
\\
After performing the angular integrations, we obtain the Eq.~\ref{eq:sigma_S} to \ref{eq:sigma_V}.

\newpage
\section{Expressions for the exchange self-energies in the case D (HF with FF+SRC calculation)}
\label{app:selfD}

\subsection{\underline{$\pi$ coupling}}

\begin{subequations}
\begin{align}
\Sigma^{\pi}_{S,N} = \frac{1}{4}\left(\frac{g_A}{2f_\pi}\right)^2 &\int{\frac{d {\bf p'}}{(2\pi)^3}}\left(\mathrm{q}^2D_{\pi}(\mathrm{q}) - 3g'_{\pi}(\mathrm{q})\right)~2\hat{M_N}(\mathrm{p'}) \nonumber \\
& +2\times \left(N\longmapsto\bar{N}\right), \\
\Sigma^{\pi}_{0,N} = \frac{1}{4}\left(\frac{g_A}{2f_\pi}\right)^2 &\int{\frac{d {\bf p'}}{(2\pi)^3}}\left(\mathrm{q}^2D_{\pi}(\mathrm{q}) - 3g'_{\pi}(\mathrm{q})\right)~2f_N(\mathrm{p'}) \nonumber \\
& +2\times \left(N\longmapsto\bar{N}\right), \\
\Sigma^{\pi}_{V,N} = -\frac{1}{4}\left(\frac{g_A}{2f_\pi}\right)^2 &\int{\frac{d {\bf p'}}{(2\pi)^3}} \left[ \left(\mathrm{q}^2D_{\pi}(\mathrm{q}) - 3h'_{\pi}(\mathrm{q})\right)  \right. \nonumber \\
& \hspace{-1.0cm} \times \left. \left( 2(\tilde{\bf p}\cdot\hat{P_N}(\mathbf{p}^\prime) ) - 4(\tilde{\bf p}\cdot\tilde{\bf q})(\hat{P_N}(\mathbf{p}^\prime)\cdot\tilde{\bf q})\right) \right. \nonumber\\ 
& \hspace{-1.0cm} \left. -2(g'_{\pi}(\mathrm{q}) - h'_{\pi}(\mathrm{q}))(\tilde{\bf p}\cdot\hat{P_N}(\mathbf{p}^\prime)) \right] \nonumber \\
& \hspace{-1.0cm} +2\times \left(N\longmapsto\bar{N}\right)
\end{align}
\end{subequations}

\subsection{\underline{$\rho^T$ coupling}}

\begin{subequations}
\begin{align}
\Sigma^{\rho^T}_{S,N} = \frac{1}{4}\left(\frac{f_{\rho}}{2M}\right)^2 & \int{\frac{d {\bf p'}}{(2\pi)^3}}\left(\mathrm{q}^2D_{\rho}(\mathrm{q}) - 3g'_{\rho}(\mathrm{q})\right)~6\hat{M}'_N \nonumber \\
& +2\times \left(N\longmapsto\bar{N}\right), \\
\Sigma^{\rho}_{0,N} = \frac{1}{4}\left(\frac{f_{\rho}}{2M}\right)^2 &\int{\frac{d {\bf p'}}{(2\pi)^3}}\left(\mathrm{q}^2D_{\rho}(\mathrm{q}) - 3g'_{\rho}(\mathrm{q})\right)2f_N(\mathrm{p'}) \nonumber \\
& +2\times \left(N\longmapsto\bar{N}\right), \\
\Sigma^{\rho}_{V,N} = -\frac{1}{4}\left(\frac{f_{\rho}}{2M}\right)^2 &\int{\frac{d {\bf p'}}{(2\pi)^3}} \left[ \left(\mathrm{q}^2D_{\rho}(\mathrm{q}) - 3h'_{\rho}(\mathrm{q})\right) \right. \nonumber \\
& \hspace{-1.0cm} \left. \times \left(2(\tilde{\bf p}\cdot\hat{P_N}(\mathbf{p}^\prime)) - 8(\tilde{\bf p}\cdot\tilde{\bf q})(\hat{P_N}(\mathbf{p}^\prime)\cdot\tilde{\bf q})\right) \right. \nonumber \\ 
& \hspace{-1.0cm} \left. -2(g'_{\rho}(\mathrm{q}) - h'_{\rho}(\mathrm{q}))(\tilde{\bf p}\cdot\hat{P_N}(\mathbf{p}^\prime)) \right] \nonumber \\
& \hspace{-1.0cm} +2\times \left(N\longmapsto\bar{N}\right)
\end{align}
\end{subequations}

\section{Derivation of the spin-isospin Landau-Migdal parameter $g^\prime$}
\label{appendix:LM}

We start from the non-relativistic reduction of the $\rho$ potential, expressed in the center of mass reference frame ($\mathbf{p}+\mathbf{p}^\prime =0$) as in Ref.~\cite{Machleidt:1987}, which provide the following contribution to the spin-isospin central interaction,
\begin{eqnarray}
\label{eq:LM interaction rho}
V_\rho(\mathbf{q})&&=-\frac{2}{3}\mathrm{q}^2 D_\rho(\mathrm{q})g_\rho^2\Big\{ 1 +2 \frac{\kappa_\rho}{2M_N} + \left(\frac{\kappa_\rho}{2M_N}\right)^2\Big\} \bm{\sigma}_1 . \bm{\sigma}_2 \nonumber\\
&&=-\frac{2}{3}V_{A,\rho}(\mathrm{q})g_\rho^2\Big\{1 +2 \frac{\kappa_\rho}{2M_N} + \left(\frac{\kappa_\rho}{2M_N}\right)^2\Big\} \bm{\sigma}_1 . \bm{\sigma}_2 \nonumber\\
\end{eqnarray}
where we have employed coupling constants consistently with our Lagrangian~\eqref{eq:L_meson}.
The spin-isospin central interaction has three components arising from the vector, vector-tensor and tensor part of the interaction.
For the $\pi$ interaction, we simply have
\begin{eqnarray}
\label{eq:LM interaction pi}
V_\pi(\mathbf{q})=-\frac{1}{3}\left(\frac{g_A}{2 f_\pi}\right)^2 V_{A,\pi}(\mathrm{q}) \bm{\sigma}_1 . \bm{\sigma}_2
\end{eqnarray}
We can now define the Landau-Migdal parameter $G'$ which governs the response in the Gamow-Teller (GT) channel and is defined as the spin-isospin channel of the central interaction,
\begin{equation}
V_{GT} = G^\prime (\bm{\sigma}_1\cdot\bm{\sigma}_2)\, .
\end{equation}
The Landau-Migdal parameter $G^\prime$ takes the contribution from the $\rho$ and the $\pi$ given in Eqs.~\eqref{eq:LM interaction rho} and \eqref{eq:LM interaction pi}, and evaluated at $\mathrm{q}=0$.

\subsection{Model A}

We have for the pure HF model (model A), from \\ Eq.~\eqref{eq:LM interaction rho},\eqref{eq:LM interaction pi}, 
\begin{equation}
G^\prime_A(\mathrm{q}) = -\frac{1}{3}\left(\frac{g_A}{2 f_\pi}\right)^2V_{A,\pi}(\mathrm{q})
-\frac{2}{3}\left(\frac{f_\rho}{2 M_N}\right)^2V_{A,\rho^T}(\mathrm{q}) \, .
\end{equation}
The dimensionless Landau-Migdal interaction $g_A^\prime(\mathrm{q})$ is defined as,
\begin{equation}
G^\prime_A(\mathrm{q}) = \left(\frac{g_A}{2 f_\pi}\right)^2
g^\prime_A(\mathrm{q}) \, ,
\end{equation}
with
\begin{equation}
g^\prime_A(\mathrm{q}) = -\frac{1}{3} V_{A,\pi}(\mathrm{q})
-\frac{2}{3} \mathcal{C}_\rho V_{A,\rho^T}(\mathrm{q}) \, ,
\end{equation}
with $\mathcal{C}_\rho = \left(f_\pi g_\rho \kappa_\rho/(g_AM_N)\right)^2$.

This expression reduces to the Landau-Migdal parameter at the long-wave length limit $\mathrm{q}\rightarrow 0$.
Since $V_{A,\alpha}(\mathrm{q}\rightarrow 0)\rightarrow 0$, we obtain for the GT Landau-Migdal parameter $g^\prime_A=0$ in the pure HF model (model A).

Note that in nuclear physics a different dimensionless parameter $\tilde{G}^\prime$ is defined as
$\tilde{G}^\prime=N_0G^\prime$, where $N_0$ is the density of state in symmetric matter and at saturation density $N_0=2M_N^* k_F/\pi^2$ ($N_0\approx 38\times10^{3}$~MeV$^2$ for $M_N^*\approx 0.7 M_N$), such that $\tilde{G}^\prime=N_0(g_A/2f_\pi)^2g^\prime\approx 1.6g^\prime$. See Ref.~\cite{GarciaRecio:1992} for more details.

For model A, we can see directly from Eqs.~\eqref{eq:LM interaction rho} and \eqref{eq:LM interaction pi} that
\begin{equation}
\label{eq:g_prime_A}
    g'_A(\mathrm{q}=0)=0.
\end{equation}

\subsection{Model B}

For model B, we replace $V_A(\mathrm{q})$ by $V_B(\mathrm{q})$ as in Eq.~\eqref{eq:caseB} for the $\rho^T$ and $\pi$ interaction, so
\begin{eqnarray}
\label{eq:g_prime_B}
G'_B &=& -\frac{1}{3}\left(\frac{g_A}{2 f_\pi}\right)^2V_{B,\pi}(\mathrm{q}=0) \nonumber\\
&&\hspace{-0.5cm}-\frac{2}{3}g_\rho^2 \Big\{ V_{A,\rho}(0) +2 \frac{\kappa_\rho}{2M_N} V_{A,\rho}(0) + \left(\frac{\kappa_\rho}{2M_N}\right)^2V_{B,\rho}(0)\Big\} \nonumber\\
&& = \frac{1}{3}\left(\frac{g_A}{2 f_\pi}\right)^2 + \frac{2}{3}\left(\frac{g_\rho \kappa_\rho}{2M_N}\right)^2 \, ,
\end{eqnarray}
and for the dimensionless Landau-Migdal parameter,
\begin{equation}
g^\prime_B = \frac{1}{3}+\frac{2}{3} \mathcal{C}_\rho \, ,
\end{equation}

\subsection{Model C}

For model C with FF, $F_\alpha(\mathrm{q}=0)=1$, so we simply get the same as in the case A, namely
\begin{equation}
\label{eq:g_prime_C}
    g'_C(\mathrm{q}=0)=0.
\end{equation}

\subsection{Model D}

Finally, for model D, we replace $V_A(\mathrm{q})$ by $V_D(\mathrm{q})$ as in Eq.~\eqref{eq:V_D} for the $\rho^T$ and $\pi$ interaction, and $V_C(\mathrm{q})$ for the rest, thus we get
\begin{eqnarray}
\label{eq:g_prime_D}
G'_D&=&-\frac{1}{3}\left(\frac{g_A}{2 f_\pi}\right)^2 V_{D,\pi}(\mathrm{0}) \nonumber\\
&&\hspace{-0.5cm}-\frac{2}{3}g_\rho^2 \Big\{ V_{C,\rho^V}(0) +2 \frac{\kappa_\rho}{2M_N} V_{C,\rho^{VT}}(0) + \left(\frac{\kappa_\rho}{2M_N}\right)^2V_{D,\rho}(0)\Big\} \nonumber\\ 
&& = \frac{1}{3}\left(\frac{g_A}{2 f_\pi}\right)^2\frac{\mathrm{q_c}^2}{\mathrm{q_c}^2+m_\pi^2} \left( \frac{\Lambda_\pi^2}{\Lambda_\pi^2+\mathrm{q_c}^2} \right)^2 \nonumber\\
&&\qquad +\frac{2}{3}\left(\frac{g_\rho \kappa_\rho}{2M_N}\right)^2 \frac{\mathrm{q_c}^2}{\mathrm{q_c}^2+m_\rho^2} \left( \frac{\Lambda_{\rho^T}^2}{\Lambda_{\rho^T}^2+\mathrm{q_c}^2} \right)^2 
\end{eqnarray}
The dimensionless Landau-Migdal parameter reads,
\begin{eqnarray}
g^\prime_D(\mathrm{q}\rightarrow 0)&=&\frac{1}{3}\frac{\mathrm{q_c}^2}{\mathrm{q_c}^2+m_\pi^2} \left( \frac{\Lambda_\pi^2}{\Lambda_\pi^2+\mathrm{q_c}^2} \right)^2 \nonumber\\
&& +\frac{2}{3} \mathcal{C}_\rho\frac{\mathrm{q_c}^2}{\mathrm{q_c}^2+m_\rho^2} \left( \frac{\Lambda_{\rho^T}^2}{\Lambda_{\rho^T}^2+\mathrm{q_c}^2} \right)^2 \, .
\end{eqnarray}
Note that if we include the SRC in all interaction channels, we would have $V_D(\mathrm{q})$ appearing for all terms in Eq.~\eqref{eq:g_prime_D} instead of $V_C(\mathrm{q})$, and the contribution from the $\rho$ and $\rho^T$ would simply add together and we would have $\kappa_\rho \longmapsto 1+\kappa_\rho$ appearing in $C_\rho$ instead.

\bibliographystyle{spphys}       
\bibliography{biblio}

\end{document}